\documentclass[%
 aip,
 amsmath,amssymb,
 reprint,%
]{revtex4-1}

\usepackage{graphicx}
\usepackage{dcolumn}
\usepackage{bm}

\usepackage[utf8]{inputenc}
\usepackage[T1]{fontenc}
\usepackage{mathptmx}
\usepackage{etoolbox}

\usepackage{multirow}
\usepackage{amsmath}
\usepackage{tabularx}
\usepackage{comment}
\usepackage{placeins}
\usepackage{caption}
\usepackage{subcaption}
\usepackage{makecell}
\usepackage{siunitx}
\usepackage{braket}

\makeatletter
\def\@email#1#2{%
 \endgroup
 \patchcmd{\titleblock@produce}
  {\frontmatter@RRAPformat}
  {\frontmatter@RRAPformat{\produce@RRAP{*#1\href{mailto:#2}{#2}}}\frontmatter@RRAPformat}
  {}{}
}%
\makeatother
\begin{document}

\preprint{AIP/123-QED}
\title[Spectroscopic characterization of the $\text{a}^\text{3} {\Pi}$ state of aluminum monofluoride]{Spectroscopic characterization of the $\text{a}^\text{3}  \boldsymbol{\Pi}$ state of aluminum monofluoride}

\author{N. Walter}
 \email{walter@fhi-berlin.mpg.de}
 \author{M. Doppelbauer}
 \author{S. Marx}
\author{J. Seifert}
\author{X. Liu}
\author{J .Pérez-Ríos}
 \affiliation{Fritz-Haber-Institut der Max-Planck-Gesellschaft, Faradayweg 4-6, 14195 Berlin, Germany}
\author{B.G. Sartakov}
 \affiliation{Prokhorov General Physics Institute, Russian Academy of Sciences, Vavilovstreet 38, 119991 Moscow, Russia}
 \author{S. Truppe}
\author{G. Meijer}
 \email{meijer@fhi-berlin.mpg.de}
 \affiliation{Fritz-Haber-Institut der Max-Planck-Gesellschaft, Faradayweg 4-6, 14195 Berlin, Germany}

\date{\today}

\begin{abstract}

Spectroscopic studies of aluminum monofluoride (AlF) have revealed its highly favorable properties for direct laser cooling. All $Q$ lines of the strong A$^1\Pi$ $\leftarrow$ X$^1\Sigma^+$ transition around 227~nm are rotationally closed and thereby suitable for the main cooling cycle.
The same holds for the narrow, spin-forbidden a$^3\Pi$ $\leftarrow$ X$^1\Sigma^+$ transition around 367~nm which has a recoil limit in the \SI{}{\micro\kelvin} range.

We here report on the spectroscopic characterization of the lowest rotational levels
in the a$^3\Pi$ state of AlF for \mbox{$v=0\,$--$\,8$} using a jet-cooled, pulsed molecular beam. An accidental AC Stark shift is observed on the a$^3\Pi_0, v=4$ $\leftarrow$ X$^1\Sigma^+, v=4$ band. 
By using time-delayed ionization for state-selective detection of the molecules in the metastable a$^3\Pi$ state at different points along the molecular beam, the radiative lifetime of the a$^3\Pi_1, v=0, J=1$ level is experimentally determined as $\tau=1.89 \pm 0.15$~ms. A laser/radio-frequency multiple resonance ionization scheme is employed to determine the hyperfine splittings in the a$^3\Pi_1, v=5$ level. The experimentally derived hyperfine parameters are compared to the outcome of quantum chemistry calculations.
A spectral line with a width of 1.27~kHz is recorded between hyperfine levels in the a$^3\Pi, v=0$ state. 
These measurements benchmark the electronic potential of the a$^3\Pi$ state and yield accurate values for the photon scattering rate and for the elements of the Franck-Condon matrix of the \mbox{a$^3\Pi$ -- X$^1\Sigma^+$} system.

\end{abstract}

\maketitle

\section{\label{sec:introduction}Introduction}

\begin{figure}[b]
		\centering
		\includegraphics[width=0.48\textwidth]{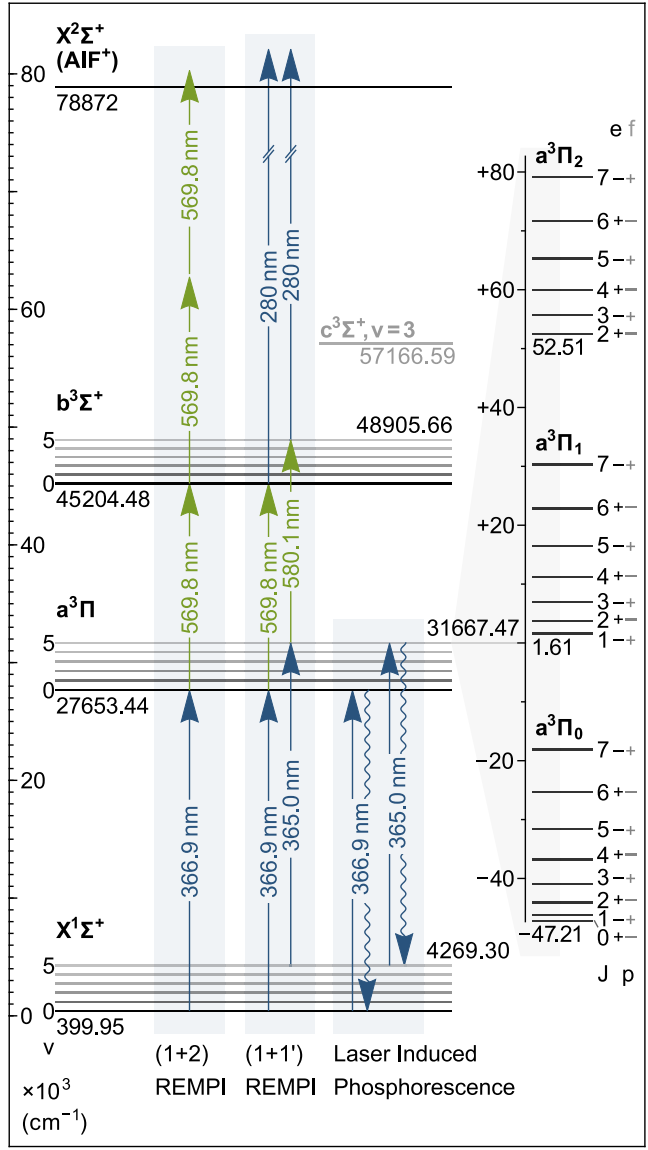}
	\caption{\label{fig:ExScheme}Scheme of the electronic, vibrational and rotational levels of AlF that are relevant in this study.
	The rotational structure of the three $\Omega$ manifolds of the a$^3\Pi$ state are shown on an expanded scale. For $\Omega=0$, the $e$ and $f$ levels are split by about 10~GHz; only the $e$ levels have been probed. 
The excitation and detection schemes that are applied are indicated. 
	In all experiments, AlF is excited on the $\Delta v=0$ bands of the a$^3\Pi$ $\leftarrow$ X$^1\Sigma^+$ transition, and thus prepared in selected vibrational and rotational levels of the metastable a$^3\Pi$ state. 
	}
\end{figure}

Ultracold atomic physics created new opportunities in precision metrology, quantum science and fundamental physics.\cite{bohn} The application of cooling techniques to molecules -- although more challenging to implement due to the more complex energy level structure of molecules -- allows full control over their motions and interactions and to study chemistry in a new regime.\cite{Heazlewood,manipulation} 
Ongoing applications include
high resolution spectroscopy to determine the possible variation of fundamental constants\cite{Chin_2009} and 
the electric dipole moment of elementary particles.\cite{Hudson2011,Baron2014,Cairncross2017,Andreev2018}
Future applications in the quantum simulation of mathematically inaccessible many body systems and quantum computing are anticipated.\cite{DeMille2002,Micheli2006,Sawant2020}

Among the various diatomic molecules that are being used in laser cooling experiments, aluminum monofluoride (AlF) has particularly favorable properties. It is strongly bound and can be efficiently produced. The photon scattering rate on the A$^1\Pi$ $\leftarrow$ X$^1\Sigma^+$ electronic transition is exceptionally large, and the vibrationally off-diagonal Einstein A coefficients are orders of magnitude smaller than the diagonal ones. Moreover, all $Q$ lines of a $^1\Pi$ $\leftarrow$ $^1\Sigma^+$ transition are rotationally closed. These favorable properties were already noted when a list of diatomic molecules was compiled for this purpose in 2004, but then it was remarked that AlF is only the most promising molecule when leaving "experimental convenience aside".\cite{dirosa} What makes laser cooling of AlF experimentally inconvenient is that the strong A$^1\Pi$ $\leftarrow$ X$^1\Sigma^+$ cooling transition is in the deep UV at around 227~nm. Moreover, both nuclei have nuclear spin, with a value of 5/2 for $^{27}$Al and 1/2 for $^{19}$F, yielding a complicated hyperfine structure. 

Recently, narrowband cw laser systems in the deep UV with a power of several hundred miliwatts have become commercially available. The hyperfine structure in the lowest rotational levels of the X$^1\Sigma^+$, A$^1\Pi$ and a$^3\Pi$ states of AlF has been experimentally determined in the required detail.\cite{alf} Efficient optical cycling has been demonstrated on various $Q$ lines of the A$^1\Pi$ $\leftarrow$ X$^1\Sigma^+$ transition.\cite{Simon} The main vibrationally off-diagonal Einstein A coefficient is experimentally found to be about 300 times smaller than the diagonal one, making laser cooling with a single repump laser feasible.
Whilst the A$^1\Pi$ $\leftarrow$ X$^1\Sigma^+$ transition allows large optical scattering forces to be generated, it has a comparatively high Doppler cooling limit of several mK. 
The a$^3\Pi$ $\leftarrow$ X$^1\Sigma^+$ transition around 367~nm is also vibrationally  highly diagonal, but its laser cooling limit is the recoil limit, which lies in the \SI[parse-numbers = false]{\micro \kelvin} range.\cite{wells}
Laser cooling can be achieved by cycling on any of the $Q_2$ lines and on the (considerably weaker) $P_1(1)$ line. 

To be able to accurately determine the Franck-Condon factors for the a$^3\Pi$ $\leftarrow$ X$^1\Sigma^+$ transition, both potential energy curves, as extracted from experimental data, need to be accurately known. For the X$^1\Sigma^+$ electronic ground state, an accurate potential has been reported.\cite{bernath2018} By contrast, the available rotationally resolved spectroscopic data on the a$^3\Pi$ state is insufficient and limited to the lowest two vibrational states.\cite{1974barrow,1978brown,1976rosenwaks} This makes the a$^3\Pi$ state the least well characterized one of the low-lying electronically excited states of AlF up to now. The lifetimes of the rotational levels in the a$^3\Pi$ state are strongly $\Omega$-, $J$- and parity-dependent.\cite{james} To determine the photon scattering rate, the radiative lifetimes of the rotational levels in the a$^3\Pi$ state need to be measured.

In this paper, we present a spectroscopic investigation of AlF in a supersonically cooled, pulsed molecular beam. Different excitation and detection schemes are applied, as indicated in Fig.~\ref{fig:ExScheme}. One-color \mbox{(1+2) REMPI} detection via the b$^3\Sigma^+, v=0$ $\leftarrow$ a$^3\Pi, v=0$ band is used to determine the lifetimes of the rotational levels in the a$^3\Pi, v=0$ state. Laser induced phosphorescence detection and two-color \mbox{(1+1$'$) REMPI} is applied to determine the energies of the lowest rotational levels up to $v=7$. The vibrational, rotational and spin-orbit coupling parameters of the a$^3\Pi$ state are extracted. All $\Lambda$-doublet transitions within the a$^3\Pi_1, v=5, J=1$ level are recorded and the hyperfine parameters are derived. Their values are compared to the outcome of {\it ab~initio} calculations. The precise knowledge of the hyperfine structure of the a$^3\Pi_1, v=5, J=1$ level is crucial for the study of the singlet-triplet doorway levels that arise from the interaction of the A$^1\Pi, v=6$ and the b$^3\Sigma^+, v=5$ states. A detailed, hyperfine resolved study of these doorway levels will be separately reported upon.\cite{nicole2}

\section{\label{sec:Setup}Experimental Setup}

\begin{figure}[b]
		\centering
		\includegraphics[width=0.48\textwidth]{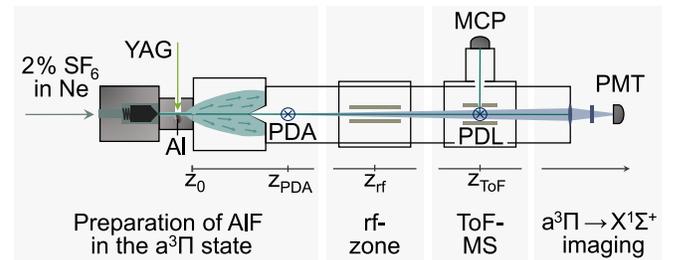}
	\caption{\label{fig:Setup} Schematic view of the modular supersonic molecular beam setup. AlF is produced in a chemical reaction of laser-ablated aluminum atoms and SF$_6$ and is excited further downstream to the metastable a$^3\Pi$ state. Subsequent excitation by laser or radio frequency radiation and detection takes place in spatially separated regions.
	}
\end{figure}

The experimental setup is shown schematically in Fig.~\ref{fig:Setup}. A pulsed solenoid valve (General Valve, Series~9) releases short gas pulses (duration about 50 \SI[parse-numbers = false]{\micro\second}, 10~Hz repetition rate) through a circular orifice into a narrow channel with a conical nozzle. The gas is composed of 2~\%~SF$_6$ seeded in a noble gas with a backing pressure of 3~bar. In the channel, the injected gas pulse passes over a rotating aluminum rod that is irradiated by a Q-switched Nd:YAG laser (1064~nm, 16~mJ pulse energy). Ablated aluminum atoms react with SF$_6$ to form AlF, and these molecules, mixed in the carrier gas, then expand from the nozzle (defined as $z=z_0=0$) into the source chamber.

In the supersonic expansion, AlF is translationally and internally cooled through collisions with the carrier gas. 
Typical translational and rotational temperatures are 
$T_{\text{transl}}\simeq 3$~K and $T_{\text{rot}} \simeq 5$~K. Due to the conditions during the production process and the relatively inefficient cooling of vibrations, the molecules are distributed over several vibrational levels of the X$^1\Sigma^+$ electronic ground state.\cite{1976rosenwaks} This is important for our study of higher vibrational levels of the metastable a$^3\Pi$ state, since only transitions of the a$^3\Pi$ $\leftarrow$ X$^1\Sigma^+$ band with $\Delta v =0$ are sufficiently strong. The population of the lowest three vibrational levels can be characterized by a vibrational temperature $T_{\text{vib}} \simeq 700$~K, but the higher vibrational levels up to $v=7$ remain populated by $\sim 1~ \%$. AlF is almost exclusively present in the electronic ground state with a very small fraction in the first electronically excited, metastable triplet state.

The supersonic expansion results in the formation of an AlF-seeded molecular beam, whose forward velocity is primarily determined by that of the carrier gas. For the lifetime measurements (see Sec.~\ref{sec:lifetime}), different noble gases are used as carrier gas in order to manipulate the beam velocity. Unless stated otherwise, all other experiments are carried out using neon, yielding a beam velocity of roughly 750~m/s. The molecular beam is collimated through a conically shaped skimmer with a 4~mm diameter opening. The coordinate $z$ (see Fig.~\ref{fig:Setup}) is defined along the axis of the molecular beam through the array of several differentially pumped vacuum chambers ($\sim 10^{-6}$~mbar). At various positions $z$ further downstream, the AlF molecules interact with laser and/or radio frequency (rf) radiation that crosses the molecular beam perpendicularly. 

At $z=z_{\text{PDA}}=25$~cm, AlF is resonantly excited from the X$^1\Sigma^+$ electronic ground state to selected ro-vibrational levels of the a$^3\Pi$ state. For this, a frequency-doubled pulsed dye amplifier (PDA), operated with pyridine~2 dye, is used. The PDA is pumped by a frequency-doubled Nd:YAG laser and injection seeded with a tunable, narrowband, cw TiSa laser. The absolute frequency of the TiSa seed-laser is measured with a wavemeter (HighFinesse~WS8, 10~MHz accuracy). The PDA contains a phase-conjugate mirror based on stimulated Brillouin scattering (SBS), and the frequency shift imposed by the SBS cell to the pulsed radiation (at the fundamental frequency) is experimentally determined to be $-1.98(5)$~GHz. The frequency-doubled laser pulses in the 365--367~nm range have about 10~mJ energy.
The spectral profile of this laser system is inferred by recording the a$^3\Pi_0,J=0$ $\leftarrow$ X$^1\Sigma^+,J=1$ transition of AlF. The hyperfine splitting of both the ground and excited levels of this transition (12~MHz and  3~MHz, respectively) is negligible compared to the laser linewidth. The measured spectral lineshape of this $P_1(1)$ line shows a near-Gaussian main peak with a full width at half maximum (FWHM) of 220~MHz, reflecting the intrinsic bandwidth of the laser system. In addition, a minor side peak with the same width, centered about 350~MHz higher in frequency, is observed. This side peak results from an adjacent longitudinal mode of the TiSa seed-laser and is taken into account in the data analysis.  

The AlF molecules in the a$^3\Pi$ state can be directly detected by monitoring the a$^3\Pi$ $\rightarrow$ X$^1\Sigma^+$ phosphorescence that occurs along the molecular beam for $z\geq z_{\text{PDA}}$. For this, an on-axis light collection system is installed at the end of the molecular beam machine and the phosphorescence is imaged on a photomultiplier tube (PMT). As the solid angle for light collection is larger when the molecules are closer to the on-axis detector, the phosphorescence signal is peaked several hundred \SI[parse-numbers = false]{\micro \second} after laser excitation. A bandpass filter centered around 367~nm in front of the PMT suppresses scattered light from the laser ablation source as well as ambient light.

Since all rotational levels in the a$^3\Pi$ state have a radiative lifetime of more than a millisecond, subsequent excitation and detection can take place in spatially separated regions, further downstream. Their arrangement is modular and adapted to the requirements of the individual experiments. 
In the rf interaction zone, transitions between opposite parity $\Lambda$-doublet components within a certain $J$ level in the a$^3\Pi$ state are driven by exposing the molecules to rf radiation in the 1--500~MHz range.  

Parity-selective detection of the molecules in the a$^3\Pi$ state is implemented by using resonant ionization via rotational levels of the b$^3\Sigma^+$ state. The AlF$^+$ cations are mass-selectively detected using a linear Wiley-McLaren time-of-flight mass spectrometer (ToF-MS). Two counter-propagating, collimated beams (a few mm diameter) of two pulsed dye lasers (PDLs) intersect the molecular beam between the extraction electrodes of the ToF-MS.

As seen in Fig.~\ref{fig:ExScheme}, two different ionization schemes are applied.
The first one is two-color resonance enhanced multiphoton ionization (\mbox{(1+1$'$) REMPI}). The radiation of the first PDL (sub-mJ pulse energy, 5~ns pulse duration, 570--580 nm) resonantly excites the molecules from the a$^3\Pi$ state to a particular ro-vibrational level in the b$^3\Sigma^+$ state. The radiative lifetime of these levels is about 190~ns.\cite{max} Within tens of ns, the molecules are ionized by a single photon of a second PDL 
(5~mJ pulse energy, 5~ns pulse duration, 280 nm). 
In the second scheme, AlF molecules in the a$^3\Pi$ state are detected via single-color \mbox{(1+2) REMPI} via the b$^3\Sigma^+$ state. For this, only the PDL in the visible, i.e. for the b$^3\Sigma^+$ $\leftarrow$ a$^3\Pi$ transition, is used but operated at higher pulse energies (5--10~mJ).

The \mbox{(1+1$'$) REMPI} scheme is used to record spectra with high resolution, since the pulse energy of the resonant excitation laser can be kept low and power broadening is suppressed. 
The main advantage of using the \mbox{(1+2) REMPI} scheme is that with only the unfocused, visible laser present the AlF$^+$ cation is the only ion produced, enabling highly sensitive detection. We observe that \mbox{(1+2) REMPI} only works for the b$^3\Sigma^+, v=0$ $\leftarrow$ a$^3\Pi, v=0$ band. We assume that in that case, the required intermediate level for the near-resonant two-photon ionization step (around 62755.5~cm$^{-1}$ above the minimum of the X$^1\Sigma^+$ state potential) is a highly excited vibrational level of the c$^3\Sigma^+$ state. The \mbox{(1+2) REMPI} scheme is exclusively used for the lifetime measurement of the a$^3\Pi, v=0$ state, as a high spectral resolution is not needed for this.
 
In order to ensure field-free excitation and ionization, the voltages on the extraction electrodes are switched on a few hundred ns after the ionization process is completed. The positively charged parent ions are then accelerated perpendicular to the plane spanned by the molecular beam and laser beams, i.e. vertically, towards two chevron mounted microchannel plates (dual MCP detector). The voltages are switched off directly after the AlF$^+$ ions have left the acceleration region, thereby spreading out higher mass ions over a wide range of arrival times. Deflection electrodes are mounted in the field-free flight path of the ToF setup, and their voltage is switched to deflect all ions that are lighter than AlF$^+$. The output of the MCP is amplified and read into a computer using a  high-speed digitizer card (NI \mbox{PXIe-5160}, maximum sample rate 2.5~GS/s).

\section{\label{sec:hamiltonian}Molecular Hamiltonian}

The Hamiltonian used to describe the energy level structure of AlF can be written in a general form as
\begin{equation}
{H}=H_{\textrm{ev}}+H_{\textrm{rot}}+H_{\textrm{fs}}+H_{\textrm{hfs}}
\label{eq:hamiltonian}
\end{equation}
and is presented and discussed in detail elsewhere.\cite{alf,max} 
In those earlier studies, the last two terms of Eq.~(\ref{eq:hamiltonian}) are used to determine the fine structure and hyperfine structure parameters for the a$^3\Pi$, A$^1\Pi$ and b$^3\Sigma^+$ states. 
Here we focus on the first two terms of Eq.~(\ref{eq:hamiltonian}), the 
electronic and vibrational part of the Hamiltonian $H_{\textrm{ev}}$
and the rotational part $H_{\textrm{rot}}$. 
Generally, the energies of the vibrational levels for a given electronic state are expressed as\cite{huberherzberg}
\begin{equation}
\begin{split}
E_{\textrm{vib}}(v)=T_{\textrm{e}}+\omega_{\textrm{e}}\left( v+\frac{1}{2}\right)-\omega_{\textrm{e}} x_{\textrm{e}}\left( v+\frac{1}{2}\right)^2+\qquad  \\
\omega_{\textrm{e}} y_{\textrm{e}}\left( v+\frac{1}{2}\right)^3 + \omega_{\textrm{e}} z_{\textrm{e}}\left( v+\frac{1}{2}\right)^4
\end{split}
\end{equation}
where $T_{\textrm{e}}$ is the minimum of the electronic potential and $\omega_{\textrm{e}}$ is the fundamental vibrational energy with the subsequent higher order correction terms $\omega_{\textrm{e}} x_{\textrm{e}}$, $\omega_{\textrm{e}} y_{\textrm{e}}$ and $\omega_{\textrm{e}} z_{\textrm{e}}$, all given in wavenumbers. A comprehensive compilation of values for these parameters for the known electronic states of AlF is given elsewhere.\cite{1974barrow,1995bernath} 

The rotational part of the Hamiltonian for the levels in the a$^3\Pi$ state is approximated by
\begin{equation}
H_{\textrm{rot}}=\tilde{A}_{v}\left(\mathbf{L}\cdot\mathbf{S}\right)+B_{v}\,(\mathbf{J}-\mathbf{L}-\mathbf{S})^2
\end{equation}
where $\tilde{A}_{v}$ is an effective electron spin-orbit coupling constant and $B_{v}$ the rotational constant for level $v$, $\mathbf{J}$ the angular momentum of the molecule, $\bf{L}$ the total orbital angular momentum of electron motion and $\mathbf{S}$ the total spin of the electrons.
We use an effective electron spin-orbit constant $\tilde{A}_{v}$, as only $e$ levels can be measured in the a$^3\Pi_0$ manifold;
there is insufficient information on the terms in the fine structure part of the Hamiltonian that contribute to the splitting between these $e$ levels and the levels in the a$^3\Pi_1$ manifold.
In a good approximation, the relation between the actual spin-orbit coupling constant $A_{v}$ and the effective constant $\tilde{A}_{v}$ used here is given by 
\begin{equation}
\label{eq:realA}
A_{v} \approx \tilde{A}_{v}+2 \lambda_{v} - (o_{v} + p_{v} + q_{v})
\end{equation}
The fine structure parameters $\lambda_{v}$ for the spin-spin interaction and $o_{v}$, $p_{v}$ and $q_{v}$ for the $\Lambda$-doubling are only known for the $v=0$ level of the a$^3\Pi$ state, \cite{alf,max} where the relation \mbox{$A_0$ $\approx$ $\tilde{A}_0$ + 590~MHz} holds. In this approximation, minor effects due to the spin-rotation interaction parameter $\gamma_{v}$ are neglected.\cite{alf} For the vibrational dependence of $\tilde{A}_{v}$ and $B_{v}$ we use the expressions
\begin{eqnarray}
\tilde{A}_{v}= \tilde{A}_{\textrm{e}} - \tilde{\zeta}_{\textrm{e}} \left( v+\frac{1}{2}\right) + \tilde{\eta}_{\textrm{e}} \left( v+\frac{1}{2}\right)^2 \\
B_{v}=B_{\textrm{e}}-\alpha_{\textrm{e}}\left( v+\frac{1}{2}\right)+\beta_{\textrm{e}}\left( v+\frac{1}{2}\right)^2
\end{eqnarray}

With $A_0/B_0 \approx85$, the a$^3\Pi$ state is close to Hund's case~(a). Both $\Lambda$ and $\Sigma$, the projections of $\mathbf{L}$ and $\mathbf{S}$ on the internuclear axis, are well defined, as is their sum $\Omega$ = $\Lambda$ + $\Sigma$. The spin-orbit interaction splits the a$^3\Pi$ state into three $\Omega$ manifolds, and the energies of the low-$J$ rotational levels are given by the eigenvalues of the rotational Hamiltonian matrix
\begin{equation}
\left ( \begin{array}{ccc}
-\tilde{A}_{v} + B_{v}(x +1) & -B_{v} \sqrt{2 \, x} & 0\\
-B_{v} \sqrt{2 \,x}  & B_{v}(x +1) & -B_{v} \sqrt{2(x-2)} \\
0 & -B_{v} \sqrt{2(x-2)} & \tilde{A}_{v}  + B_{v}(x -3) + c_{v}
\end{array} \right )
\label{eq:pertmatrix}
\end{equation}
with $x = J(J+1)$. The correction term~$c_{v}= 4 \lambda_{v} - (o_{v} + p_{v} + q_{v})$
is needed because of the use of $\tilde{A}_{v}$ instead of $A_{v}$,
and is kept fixed at $c_{v} = c_{\textrm{0}} = 2 \lambda_{\textrm{0}} + 590$~MHz.
The first, second and third row and column have the wavefunctions for $\Omega=0$, 1 and 2 as basis. $\Lambda$-doubling in the $\Omega=1$ and 2 manifolds is neglected. Commonly, an additional offset term $B_{v}$ is added to the diagonal elements of the matrix,\cite{Brown1979b,COFiled} but here we follow the convention used in earlier studies on AlF by Brown
~{\it et}~{\it al.}\cite{1978brown} 
Note that in this case also the term $-4/3 \, \lambda_{v}$ is absorbed in the value of $T_{\textrm{e}}$.

The X$^1\Sigma^+$ and the b$^3\Sigma^+$ states are best described by Hund's case~(b). In this case, the total angular momentum $\mathbf{J}$ is the vectorial sum of the end-over-end rotation $\mathbf{N}$ and the total spin of the electrons $\mathbf{S}$; for the X$^1\Sigma^+$ state, $\mathbf{J}=\mathbf{N}$. The expression for the unperturbed rotational energy is given by $B_{v} \,\, N(N+1)$. In the present study, the b$^3\Sigma^+$ state is merely used to enable parity- and rotational-level-selective detection of molecules in the a$^3\Pi$ state. The further splitting of the rotational levels due to the fine structure and hyperfine structure parts of the Hamiltonian, for which the explicit expressions are given elsewhere,\cite{max} is therefore not relevant and is not discussed here.

\section{\label{sec:Rot}Rotationally resolved spectroscopy of
the $\text{a}^\text{3} \boldsymbol{\Pi} \text{, v}$ $\leftarrow$ 
$\text{X}^\text{1}  \boldsymbol{\Sigma^+} \text{, v}$ bands}

\begin{figure}[b]
		\centering
		\includegraphics[width=0.48\textwidth]{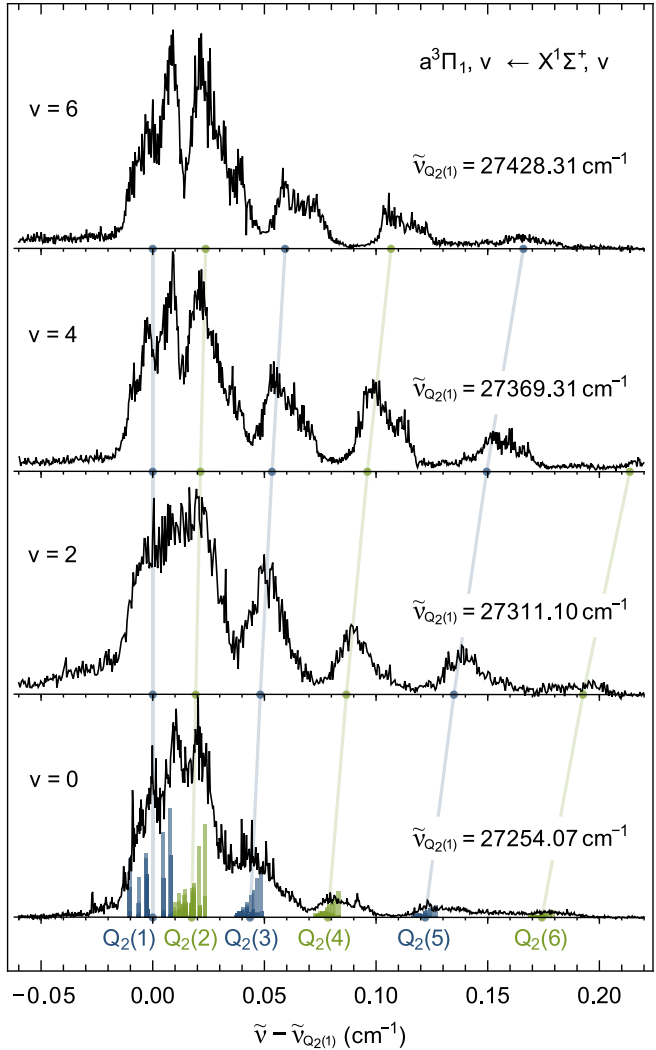}
	\caption{\label{fig:QBranch} Measured $Q_2$ lines of the a$^3\Pi_1, v$ $\leftarrow$ X$^1\Sigma^+, v$ bands for $v=0$, 2, 4 and 6. The molecules in the metastable state are detected using two-color \mbox{(1+1$'$) REMPI} via the b$^3\Sigma^+, v$ state. The spectra are shifted such that the centers of the $Q_2(1)$ lines align. The calculated hyperfine structure of the $Q_2$ lines is given as a stick spectrum.}
\end{figure}

\begin{table}
\caption{\label{tab:arotpar} 
Fitted spectroscopic parameters for the a$^3\Pi$ state from this work, compared to values reported earlier\cite{1976kopp} and to the values for the X$^1\Sigma^+$ state.\cite{1995bernath} All values are given in cm$^{-1}$.}
 \begin{ruledtabular}
 
\begin{tabular}{c|lll}
Parameter &
a$^3\Pi$, this work  & 
a$^3\Pi$, Ref.~\onlinecite{1976kopp} &
X$^1\Sigma^+$, Ref.~\onlinecite{1995bernath}\\ [0.5ex]
\hline
$T_{\textrm{e}}$ & 27239.4529(53) & 27239.4(1) & 0 \\
$\omega_{\textrm{e}}$ & 830.2807(11) & 830.3(3) & 802.32447(11) \\
$\omega_{\textrm{e}} x_{\textrm{e}}$ & 4.64364(58) & 4.6(2) & 4.849915(44) \\
$\omega_{\textrm{e}} y_{\textrm{e}}$ & 0.01230(12) & - & 0.0195738(68) \\
$\omega_{\textrm{e}} z_{\textrm{e}}$ & $-$0.0001119(78) & - & $-$0.00003407(35) \\
$\tilde{A}_{\textrm{e}}$ & 47.3475(17) & - & - \\
$\tilde{\zeta}_{\textrm{e}}$ & $-$0.07391(28) & - & - \\
$\tilde{\eta}_{\textrm{e}}$ & $-$0.000695(48) & - & -\\
$B_{\textrm{e}}$ & 0.557376(48) & 0.55718(3) & 0.552480208(65) \\
$\alpha_{\textrm{e}}$ & 0.004798(30) & 0.00468(5) & 0.004984261(44) \\
$\beta_{\textrm{e}}$ & 0.0000258(40) & - & 0.000017215(95) \\
\end{tabular}
\end{ruledtabular}
\end{table}

Here we present our spectroscopic investigation of the rotational structure of the metastable a$^3\Pi$ state up to $v=7$. The spin-forbidden a$^3\Pi$ $\leftarrow$ X$^1\Sigma^+$ transition gets its intensity mainly from spin-orbit coupling of the a$^3\Pi$ state with a $^1\Pi$ state and, to a lesser extent, from spin-orbit coupling with a $^1\Sigma^+$ state.\cite{alf} 
The $\Delta \Omega =0$ selection rule for these couplings results in transitions to the a$^3\Pi_1$ manifold being the strongest, about a factor 20 stronger than those to the $e$ levels of the a$^3\Pi_0$ manifold. 
Transitions to the a$^3\Pi_0$, $f$ levels and to the a$^3\Pi_2$ manifold are again about two orders of magnitude weaker.
For more details on the spin-orbit coupling, see Sec.~\ref{sec:lifetime}.
Since the Franck-Condon matrix between the a$^3\Pi$ and X$^1\Sigma^+$ states is highly diagonal, only $\Delta v=0$ bands are used in the experiments. 

The characteristics of the TiSa-seeded PDA that is used for a$^3\Pi$ $\leftarrow$ X$^1\Sigma^+$ excitation are given in Sec.~\ref{sec:Setup}.
The setup, shown in Fig.~\ref{fig:Setup}, is arranged such that the a$^3\Pi$ $\rightarrow$ X$^1\Sigma^+$ phosphorescence is detected at $z=65$~cm. This technique is used for a preliminary search of the energy levels of higher vibrational states of the a$^3\Pi$ state. In between excitation and phosphorescence detection, at $z=45$~cm, a ToF-MS is installed for \mbox{(1+1$'$) REMPI}. This technique is more sensitive and allows the accurate determination of selected ro-vibrational transition frequencies. The frequencies of the $Q_2$ and the  \mbox{$R_2(0)$ lines} are measured for $v\leq7$ and the transition frequencies of the \mbox{$P_1(1)$ lines} lines are measured for $v\leq5$.
The pulse energy of the PDA (around 10~mJ) is sufficient to saturate the transitions to the a$^3\Pi_1$ manifold and to efficiently excite molecules to the $e$ levels in the a$^3\Pi_0$ manifold. The signal intensity is mainly limited by the thermal population of the molecules in the initial vibrational level $v$ in the X$^1\Sigma^+$ ground state. 

In Fig.~\ref{fig:QBranch}, the measured $Q_2$ branches are shown for selected diagonal bands of the a$^3\Pi_1, v$ $\leftarrow$ X$^1\Sigma^+, v$ transition. 
It is evident from these spectra that the value for $\Delta B_v$, 
defined as the difference between the rotational constants in the a$^3\Pi, v$ and X$^1\Sigma^+, v$ states, increases with $v$. The hyperfine structure cannot be resolved in these measurements and leads to a partial overlap of the $Q_2(1)$ and $Q_2(2)$ lines. The center frequency of the $Q_2(1)$ lines is determined from the measured center frequency of the spectrally isolated $R_2(0)$ lines, using the known rotational energy structure of the electronic ground state.\cite{1995bernath} The $Q_2(1)$  and $R_2(0)$ lines reach different parity components of the a$^3\Pi_1, J=1$ level, but their splitting can be neglected (see Sec.~\ref{sec:hyperfinev5}). The spectra are shifted such that the centers of the $Q_2(1)$ lines are aligned at $\tilde{\nu}=0$. The dots indicate the transition frequencies of the $Q_2(J)$ lines. The calculated hyperfine structure of the transitions to the a$^3\Pi_1,v=0,J$  levels is given as a stick spectrum underneath, with their center of gravity aligned to the observed line centres. Note that the discrepancy that is evident between the line positions and the measured spectra is due to the side peak of the spectral profile of the PDA.  

In total, 30 clearly separated lines are measured and their center frequencies are determined using Gaussian fits and listed in Table~\ref{tab:RotTable} in Appendix~\ref{ap:RotLineList}, together with the deviations between the measured and calculated frequencies.
The spectroscopic parameters of the a$^3\Pi$ state deduced from a fit to all the measured frequencies are given in Table~\ref{tab:arotpar}. The standard deviation of the fit is about 40~MHz.   

The effective electron spin-orbit coupling constant $\tilde{A}_{v}$ increases with $v$ while the rotational constant $B_{v}$ decreases with $v$.
Using the expression for $A_0$ given in Eq.~(\ref{eq:realA}), we find a deviation of 11~MHz with the independently determined value.\cite{max} The value obtained for $B_0$ agrees to within 4~MHz with the value found earlier.\cite{alf} The energy of the center of gravity of the $J=1$ level in the a$^3\Pi_1, v=0$ state relative to that of the $N=0$ level in the X$^1\Sigma^+, v=0$ state is consistent within 40~MHz to the value given earlier.\cite{max}

\begin{table}
\caption{\label{tab:EMO} 
Parameters of the empirical EMO potential for the a$^3\Pi$ state of AlF.
The dissociation energy and the equilibrium distance are kept fixed at 
$D_{\textrm{e}}=28324.6$~cm$^{-1}$
and $r_{\textrm{e}}= 1.64708$~\AA.
The dimensionless root mean square deviation is 0.58. All values in~\AA$^{-1}$.
}
 \begin{ruledtabular}
\begin{tabular}{lll}
  $\beta_0= 1.995686(69)$& $\beta_3= -21.6(4.3)$ &   $\beta_6= 1006(210)$ \\
   $\beta_1=0.0863(43)$&$\beta_4=233(43)$&\\
$\beta_2= 0.26(12)$   &$\beta_5= -815(160)$&\\
\end{tabular}

\end{ruledtabular}
\end{table}

The experimentally derived spectroscopic constants are converted to an analytical potential using the least-square fitting procedure available through Le~Roy's program.\cite{dPotFit,betaFit,RKR} 
In particular, we obtain a reliable RKR potential in a point-wise fashion utilizing the Dunham coefficients. This potential is then fitted to a given functional form through the betaFIT program of Le~Roy,\cite{betaFit} which in this case is the expanded Morse oscillator (EMO) potential function as 
\begin{equation}
\label{eq7}
V_{\text{EMO}}(r)=D_{\textrm{e}}\left(1-e^{-\beta(r)(r-r_{\textrm{e}})} \right)^2
\end{equation}
where $r_{\textrm{e}}$ represents the equilibrium distance, $D_{\textrm{e}}$ the dissociation energy and
\begin{equation}
\label{eq8}
\beta(r)=\sum_{\text{i}=0}^{N_{\beta}}\beta_\text{i} [y_{\text{ref}}^q(r)]^\text{i}
\end{equation}
Here, $N_\beta$ accounts for the number of anharmonic terms included in the potential and
\begin{equation}
\label{eq9}
y_{\text{ref}}^q(r)=\frac{r^q-r_{\text{ref}}^q}{r^q+r^{q}_{\text{ref}}}
\end{equation}
represents a normalized interatomic distance designed to enhance the precision of the fitting procedure. 
The results of the fitting procedure for the a$^3\Pi$ state of AlF are summarized in Table~\ref{tab:EMO}. 
In the fit, the reference distance $r_\text{ref}=1.45$~\AA~and $q=2$ are used to fit the point-wise RKR potential into an EMO potential.
The equilibrium distance $r_{\textrm{e}}$ and the dissociation energy $D_{\textrm{e}}$ are maintained as fixed parameters. The last one is calculated using the results of this study, combined with the recent and most accurate measurement of the dissociation energy of the X$^1\Sigma^+$ state.\cite{bernath2018}

\section{\label{sec:AC}AC Stark shifts on the $\text{a}^\text{3} \boldsymbol{\Pi_0} \text{, v = 4}$ $\leftarrow$ 
$\text{X}^\text{1}  \boldsymbol{\Sigma^+} \text{, v = 4}$
band
}

\begin{figure}[b]
		\centering
		\includegraphics[width=0.48\textwidth]{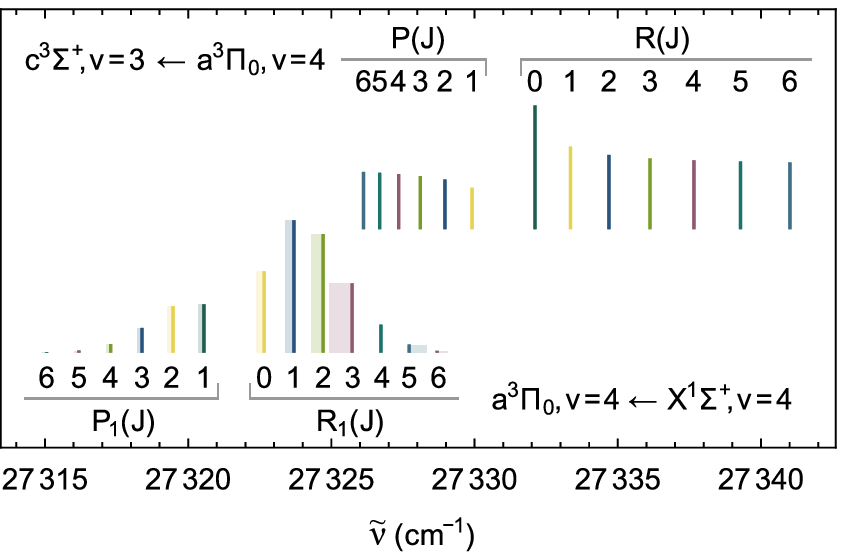}
	\caption{\label{fig:AC1}Calculated line positions of the a$^3\Pi_0, v=4$ $\leftarrow$ X$^1\Sigma^+,v=4$ and c$^3\Sigma^+,v=3$ $\leftarrow$ a$^3\Pi_0, v=4$ bands. Transitions that have a $J$ level in the a$^3\Pi_0, v=4$ state in common have the same color. The heights of the lines in the upper spectrum reflect the H\"onl-London factors, whereas in the lower spectrum these reflect the expected line intensities for $T_{\text{rot}}=5$~K. The expected (relative) magnitude and sign of the AC Stark shift is indicated by the shaded boxes in the lower spectrum. }
\end{figure}

\begin{figure}[b]
		\centering
		\includegraphics[width=0.48\textwidth]{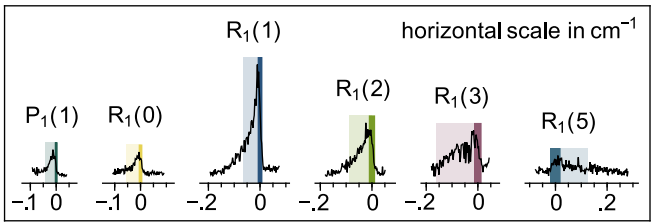}
	\caption{\label{fig:AC2} Measured lineshapes together with their expected shading. The widths of the solid vertical bars are the spans of the hyperfine structure of the transitions.}
\end{figure}

While recording the $P_1(1)$ line of the a$^3\Pi_0,v=4$ $\leftarrow$ X$^1\Sigma^+,v=4$ band, we noticed that the expected 220~MHz wide single line is asymmetrically shaped. The line is broadened to about 500~MHz and shaded to lower frequencies such that the center frequency appears about 200~MHz lower than anticipated. For that reason, this line is left out of the fitting of the spectroscopic parameters of the a$^3\Pi$ state and is not listed in Table~\ref{tab:RotTable}. Other rotational lines of the same band show distorted, asymmetric lineshapes as well, broadened significantly beyond the expected span of the hyperfine structure. As explained in the following, these observations are well accounted for by AC Stark shifts due to near-resonant rotational lines of the much more intense c$^3\Sigma^+, v=3$ $\leftarrow$ a$^3\Pi_0, v=4$ band. 

In order to efficiently induce the weak transitions from the X$^1\Sigma^+$ state to the a$^3\Pi_0$ manifold, laser pulses of about 10~mJ energy, 5~ns duration and with a bandwidth of 220~MHz are used. The approximately circular collimated beam has a diameter of several~mm. In the presence of other, strong transitions that are coincidentally nearby and involve the same $J$ level in the a$^3\Pi_0$ state, the high electric field $\mathcal{E}(\tilde{\nu}_\textrm{l})$ associated with the laser pulse at frequency $\tilde{\nu}_\textrm{l}$ causes a shift of this $J$ level. 
The magnitude of the energy shift, $\delta E(J)$, is given by\cite{zare}
\begin{equation}
\label{eq:StarkShiftGeneral}
\delta E(J) \propto \sum_i \frac{|\mu_\textrm{i}\mathcal{E}(\tilde{\nu}_\textrm{l})|^2}{h(\tilde{\nu}_\textrm{l} - \tilde{\nu}_\textrm{i})}
\end{equation}
with the transition frequencies $\tilde{\nu}_\textrm{i}$, the transition dipole moments $\mu_\textrm{i}$ and the Planck's constant $h$.
It is evident from Eq.~(\ref{eq:StarkShiftGeneral}) that only levels with $\tilde{\nu}_\textrm{i} \approx \tilde{\nu}_\textrm{l}$ contribute significantly to the total shift $\delta E(J)$.

The origin of the a$^3\Pi_0, v=4$ $\leftarrow$ X$^1\Sigma^+,v=4$ band is about 10~cm$^{-1}$ below the origin of the c$^3\Sigma^+,v=3$ $\leftarrow$ a$^3\Pi_0, v=4$ band, as schematically shown in Fig.~\ref{fig:AC1}. The latter transition is spin-allowed, and even though $\Delta v=-1$, the transition dipole moment is estimated to be about three orders of magnitude larger than for the spin-forbidden a$^3\Pi_0, v=4$ $\leftarrow$ X$^1\Sigma^+,v=4$ transition. When the laser is tuned to resonance with the $P_1(1)$ line of the a$^3\Pi_0, v=4$ $\leftarrow$ X$^1\Sigma^+,v=4$ band, the near-resonance of the laser frequency with the $R(0)$ transition of the c$^3\Sigma^+,v=3$ $\leftarrow$ a$^3\Pi_0, v=4$ band shifts the $J=0$ level in the a$^3\Pi_0, v=4$ to lower energy. For levels with $J>0$ in the a$^3\Pi_0, v=4$ state, both the $P(J)$ and the $R(J)$ lines of the c$^3\Sigma^+,v=3$ $\leftarrow$ a$^3\Pi_0, v=4$ band contribute to its overall shift. In Fig.~\ref{fig:AC1}, transitions that have a $J$ level in the a$^3\Pi_0, v=4$ state in common are indicated by the same color. Their relative contribution scales with the normalized H\"onl-London factor for a $^3\Sigma^+$ $\leftarrow$ $^3\Pi_0$ transition, indicated in Fig.~\ref{fig:AC1} by the height of the lines in the c$^3\Sigma^+,v=3$ $\leftarrow$ a$^3\Pi_0, v=4$ spectrum. 
The frequency shift $\delta \tilde{\nu}(J)$ expected for the $P_1(J+1)$ or $R_1(J-1)$ lines of the a$^3\Pi_0, v=4$ $\leftarrow$ X$^1\Sigma^+,v=4$ band that reach the level $J$ in the a$^3\Pi_0, v=4$ state is given by 
\begin{equation}
\delta \tilde{\nu}(J) \propto \frac{I}{2J+1} \left(
\frac{J}{\tilde{\nu} - \tilde{\nu}_{P(J)}} + \frac{J+1}{\tilde{\nu} - \tilde{\nu}_{R(J)}}
\right)
\end{equation}
Here, $I$ is the intensity of the laser at the (unperturbed) frequency $\tilde{\nu}$ for the $P_1(J+1)$ or $R_1(J-1)$ lines of the a$^3\Pi_0, v=4$ $\leftarrow$ X$^1\Sigma^+,v=4$ band, respectively. The frequencies $\tilde{\nu}_{P(J)}$ and $\tilde{\nu}_{R(J)}$ are those of the $P(J)$ and $R(J)$ lines of the c$^3\Sigma^+,v=3$ $\leftarrow$ a$^3\Pi_0, v=4$ band, i.e. from a$^3\Pi_0, v=4, J$ to c$^3\Sigma^+, v=3, N=J-1$ and c$^3\Sigma^+, v=3, N=J+1$, respectively.

In the experiment, the AlF molecules in the beam interact with the pulsed radiation of the PDA and experience an intensity that varies over time. In addition, as the molecules can be considered stationary while they interact with the laser pulse, they experience an intensity that depends on their position in the laser beam. This explains why we do not observe an overall spectral shift of the $P_1(1)$ line, but rather an asymmetric broadening and shading towards lower frequencies. The temporal and spatial profile of the laser beam is the same throughout the spectral range of the a$^3\Pi_0, v=4$ $\leftarrow$ X$^1\Sigma^+,v=4$ band and its effect as imprinted in the spectral shape of the $P_1(1)$ line can be used to model all the other lines in the spectrum. The $P_1(1)$ line is particularly suited for this because it only experiences an AC Stark shift due to the $R(0)$ line of the c$^3\Sigma^+,v=3$ $\leftarrow$ a$^3\Pi_0, v=4$ band and, as mentioned earlier, the $P_1(1)$ line is intrinsically very narrow. 

To model the other $P_1(J+1)$ and $R_1(J-1)$ lines of the a$^3\Pi_0, v=4$ $\leftarrow$ X$^1\Sigma^+,v=4$ band that reach levels with $J>0$ in the a$^3\Pi_0, v=4$ state, we take the observed spectral structure of the $P_1(1)$ line, stretched on the frequency axis with the calculated ratio of $\delta \tilde{\nu}(J)$/$\delta \tilde{\nu}(0)$. In the stick-spectrum in Fig.~\ref{fig:AC1}, as well as for the measured lines in Fig.~\ref{fig:AC2}, this ratio is indicated by the shaded area.
The calculated ratio $\delta \tilde{\nu}(J)$/$\delta \tilde{\nu}(0)$ decreases for increasing values of $J$ in the $P_1$ branch, whereas in the $R_1$ branch it increases from $J=1$ to $J=4$, peaks for $J=5$ and changes sign at $J=6$ and decreases again for increasing $J$.

The large value at $J=5$ is due to the very near coincidence of the $R_1(4)$ line of the a$^3\Pi_0, v=4$ $\leftarrow$ X$^1\Sigma^+,v=4$ band with the $P(5)$ line of the c$^3\Sigma^+,v=3$ $\leftarrow$ a$^3\Pi_0, v=4$ band, broadening the $R_1(4)$ line to an extent such that we have not observed it. Levels with $J>0$ in the a$^3\Pi_0, v=4$ state have an additional contribution to the width of the observed spectral lines due to the hyperfine structure in the a$^3\Pi_0, v=4$ state that spans approximately $J \cdot \,$200~MHz. This simple model represents the observed spectral shapes very well, as is shown for the six rotational lines in Fig.~\ref{fig:AC2}.

The origin of the C$^1\Sigma^+,v=0$ $\leftarrow$ a$^3\Pi_0, v=4$ band is only about 1.5~cm$^{-1}$ below the origin of the a$^3\Pi_0, v=4$ $\leftarrow$ X$^1\Sigma^+,v=4$ band. The transition dipole-moment of this spin-forbidden, $\Delta v=-4$ band is very small and it is not expected to cause a visible AC Stark shift of the levels in the a$^3\Pi_0, v=4$ state; the magnitude and direction of the shading of the spectral lines would be quite different from the observed one. 
There is an an accidental overlap ($< 0.10$~cm$^{-1}$) between the $R_1(0)$ line of the a$^3\Pi_0, v=4$ $\leftarrow$ X$^1\Sigma^+,v=4$ band and the $R(1)$ line of the C$^1\Sigma^+,v=0$ $\leftarrow$ a$^3\Pi_0, v=4$ band. 
This overlap probably causes the $S(0)$ two-photon transition from the X$^1\Sigma^+, v=4, N=0$ level to the C$^1\Sigma^+, v=0, N=2$ level, followed by one-photon ionization, to be quite strong, and likely explains why the $R_1(0)$ line of the a$^3\Pi_0, v=4$ $\leftarrow$ X$^1\Sigma^+,v=4$ band (shown in Fig.~\ref{fig:AC2}) is considerably weaker than expected.

\section{\label{sec:lifetime}Radiative lifetimes in the $\text{a}^\text{3} \boldsymbol{\Pi} \text{, v = 0}$ state}

\begin{figure}[b]
		\centering
		\includegraphics[width=0.48\textwidth]{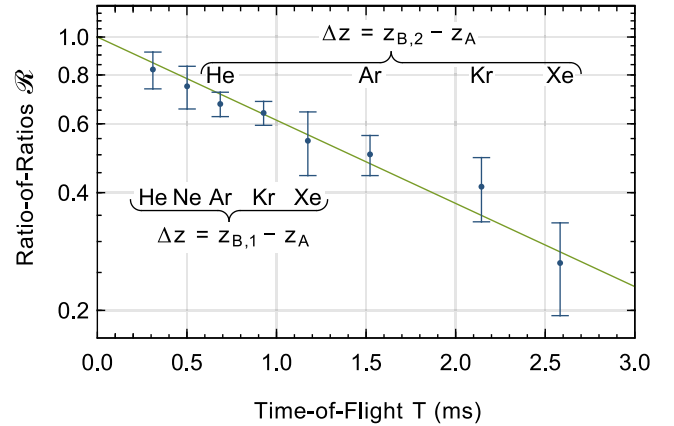}
	\caption{\label{fig:LifetimeExp}Measured values for  the "ratio-of-ratios" $\mathcal{R}(T)$ as a function of the time-of-flight $T$, shown on a semi-log plot.
	A fit to the data gives the characteristic lifetime \mbox{$\tau_{\text{exp}}$ = 2.04 $\pm$ 0.16~ms.}
	}
\end{figure}

\begin{figure}[b]
		\centering
		\includegraphics[width=0.48\textwidth]{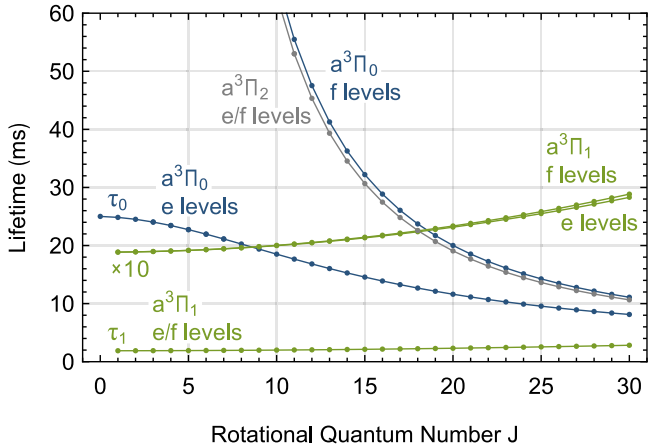}
	\caption{\label{fig:LifetimeCalc}Calculated lifetimes of the rotational levels in the a$^3\Pi, v=0$ state of AlF. The absolute scale is set by the lifetime of the a$^3\Pi_1, v=0,J=1,e$ level, experimentally determined as $\tau_1=1.89 \pm 0.15$~ms. The lifetimes of the rotational levels in the $\Omega=1$ manifold are also shown on an expanded ($\times$10) scale.
	}
\end{figure}

As mentioned before, the a$^3\Pi$ $\leftarrow$ X$^1\Sigma^+$ transition becomes weakly allowed due to spin-orbit coupling of the a$^3\Pi$ state with a $^1\Pi$ state and, to a lesser extent, with a $^1\Sigma^+$ state. The $\Delta \Omega =0$ selection rule for these couplings makes transitions to the a$^3\Pi_1$ manifold about a factor 20 stronger than those to the a$^3\Pi_0$ manifold. Both the $e$  and $f$  levels of the a$^3\Pi_1$ manifold can be reached from the X$^1\Sigma^+$ state, and the transitions to the lowest $J$ levels are the strongest. As only the $e$ levels of the a$^3\Pi_0$ manifold can interact with a $^1\Sigma^+$ state, only these can be reached from the electronic ground state for low values of $J$. For higher values of $J$, the $\Omega=1$ character in the wavefunctions of the levels in the a$^3\Pi_2$ and a$^3\Pi_0$ manifolds increases, see Table~\ref{tab:cfactors}. Consequently, transitions from the electronic ground state to higher rotational levels become gradually more allowed. The spin-orbit coupling mechanism makes the lifetimes of the rotational levels in the a$^3\Pi$ state strongly $\Omega$-, $J$- and $e/f$-dependent.\cite{james} 
The lifetime of the a$^3\Pi_1, J=1$ level is the shortest.
The low-$J$, $e$   and $f$  levels of the a$^3\Pi_2$ manifold and the low-$J$, $f$  levels of the a$^3\Pi_0$ manifold live up to three orders of magnitude longer; the 
a$^3\Pi_0, J=0,f$ level cannot decay to the electronic ground state via an electric dipole allowed transition (see Appendix~\ref{ap:lifetimeaDerivation}, Table~\ref{tab:LifetimeCalc}). 

Measuring the lifetime of metastable states of molecules is challenging, as neutral molecules can generally not be kept long enough in the observation volume. As discussed above, the lifetimes of electronically excited, metastable states generally strongly depend on the exact (rotational) quantum level, requiring state-selective preparation or detection techniques. An accurate measurement of the radiative lifetime of selected rotational levels of CO in the \mbox{a$^3\Pi, v=0$} state could only be performed when these molecules were state-selectively prepared and subsequently decelerated and electrostatically trapped.\cite{gilijamse2007} The thus determined lifetime is in excellent agreement with the result of a recent relativistic, {\it ab~initio} calculation.\cite{Mosyagin} Prior to these measurements with trapped CO molecules, the most accurate results were obtained from molecular beam experiments, employing two spatially separated LIF detection zones.\cite{Jongma1997} These experiments determined a lifetime that is 30\% too large, exemplifying the inherent challenges of such measurements. For the here presented lifetime measurement of the a$^3\Pi, v=0$ state of AlF, we use an improved method that reduces systematic errors. In this method, the signal from the molecules in the $\Omega=1$ manifold is calibrated against that of the molecules in the $\Omega=0$ manifold.

The molecules are excited either on the $R_2(0)$ or on the $R_1(0)$ line to the $J=1, e$ level of the a$^3\Pi_1,v=0$ or a$^3\Pi_0,v=0$ manifold. The measured, spectrally integrated intensities of these two lines are given by $I_{R_2(0)}$ and $I_{R_1(0)}$.
We state-selectively detect the population in the metastable state in two spatially separated ToF mass spectrometers and measure the ratio of the two ion-signals as a function of the time-of-flight $T$ between the two detection regions. 
The value of $T$ is varied by using different carrier gases and by changing the distance between the two mass spectrometers. 
The first detection zone~(A) is mounted at $z_\text{A}=z_\text{PDA}+5.5$~cm, and the second one~(B) either at $z_\text{B,1}=z_\text{PDA}+47.0$~cm or at $z_\text{B,2}=z_\text{PDA}+97.0$~cm.
The population in the a$^3\Pi_{\Omega}, v=0,J=1,e$ level at $z=z_\text{A}$ is referred to as $N_{\Omega,\text{A}}(T=0)$ and the measured ion signal is referred to as $S_{\Omega,\text{A}}(T=0)$. The population in this same level at $z=z_\text{B}$ is referred to as $N_{\Omega,\text{B}}(T)$ and the corresponding signal intensity as $S_{\Omega,\text{B}}(T)$.
The lifetime of the a$^3\Pi_{\Omega}, v=0,J=1,e$ level is labelled as $\tau_{\Omega}$. Using these expressions, we get for the ratio of the ion signals measured in the two detection zones
\begin{equation}
\label{equation:lifetime1}
     \frac{S_{\Omega,\text{B}}(T)}{S_{\Omega,\text{A}}(T=0)}=
      \frac{\alpha_{\Omega,\text{B}}}
      {\alpha_{\Omega,\text{A}}}
        \frac{N_{\Omega,\text{B}}(T)}
      {N_{\Omega,\text{A}}(T=0)}
      =
     \frac{\alpha_{\Omega,\text{B}}}{\alpha_{\Omega,\text{A}}}e^{-T/\tau_{\Omega}}
\end{equation}
with $\alpha_{\Omega,\text{A}}$ and $\alpha_{\Omega,\text{B}}$ the detection efficiencies of zone~(A) and~(B). When the PDA and the ionization laser are aligned in the same way throughout the whole measurement, we can assume that  $\alpha_{0,\text{B}}/\alpha_{0,\text{A}}=\alpha_{1,\text{B}}/\alpha_{1,\text{A}}$. Then, the "ratio-of-ratios" $\mathcal{R}(T)$ is given by
\begin{equation}
\label{equation:lifetime2}
   \mathcal{R}(T)=
   \left.
   \frac{S_{1,\text{B}}(T)}{S_{1,\text{A}}(T=0)} \middle/
   \frac{S_{0,\text{B}}(T)}{S_{0,\text{A}}(T=0)}
   \right.=e^{-T/\tau_{\text{exp}}}
\end{equation}   
where
\begin{equation}
     ~\frac{1}{\tau_{\text{exp}}} = \frac{1}{\tau_1}-\frac{1}{\tau_0}
\end{equation}
To determine the absolute values of $\tau_1$ and $\tau_0$ from $\tau_{\text{exp}}$, the ratio of $\tau_1$ and $\tau_0$ needs to be known. The latter can be deduced from $I_{R_2(0)} / I_{R_1(0)}$.

In the experiment, the vertical spread of the molecular beam is confined by a 2~mm slit, placed just in front of the intersection with the 367~nm laser. The size of the PDA preparation beam is constrained by an iris aperture, transmitting 3~mJ of pulse energy in a 3~mm diameter beam. AlF is ionized using \mbox{(1+2) REMPI} via the b$^3\Sigma^+$ state. The ionization laser beam is defined with 10~mm diameter apertures to avoid striking the ToF electrodes. This laser passes first through zone~(B) and then through zone~(A); the pulse energy is about 7~mJ in zone~(B) and 5~mJ in zone~(A). The alignment of the ionization laser stays the same throughout; switching between the two detection regions is achieved by changing the timing of the laser pulse. The values for the "ratio-of-ratios" $\mathcal{R}(T)$ are recorded as a function of $T$ by using helium, neon, argon, krypton and xenon as carrier gases and the two different ToF-positions $z_\text{B,1}$ and $z_\text{B,2}$. The results of the measurements are shown in Fig.~\ref{fig:LifetimeExp}. The error bars show the standard deviation from a series of measurements for every data point. An exponential function according to Eq.~(\ref{equation:lifetime2}) fitted to the data yields a value of $\tau_{\text{exp}} = 2.04 \pm 0.16$~ms. 

The  intensity ratio $I_{R_2(0)}/I_{R_1(0)}$ is measured as $20.9 \pm 1.0$. The different intensities of these lines is due to the different degree of coupling of the a$^3\Pi_{\Omega}, v=0,J=1, e$ levels to the X$^1\Sigma^+, v=0,N=0$ level. This coupling is due to transition dipole moments obtained from mixing with pure singlet states, either $^1\Pi$ or $^1\Sigma^+$ states.
These separate contributions can either interfere constructively or destructively. The observation that the $P_1$ lines are considerably weaker than the $R_1$ lines\cite{alf} is only consistent with the case of destructive interference. From the analysis given in  Appendix~\ref{ap:lifetimeaDerivation} it is derived that for destructive interference $\tau_0$/$\tau_1 = 13.1 \pm 1.1$.
From this we conclude that the lifetime of the a$^3\Pi_1, v=0,J=1,e$ level is $\tau_1 = 1.89 \pm 0.15$~ms and the lifetime of the a$^3\Pi_0, v=0,J=1,e$ level is $\tau_0 = 24.8 \pm 2.9$~ms. 

In Fig.~\ref{fig:LifetimeCalc}, the calculated lifetimes for the rotational levels in the a$^3\Pi, v=0$ state of AlF are shown. For several of the lowest $J$ levels, these lifetimes are listed in Table~\ref{tab:LifetimeCalc}. These values are calculated using the known spectroscopic constants in the a$^3\Pi, v=0$ state, assuming the lifetimes are solely determined by electric dipole allowed transitions and taking both the $J$-dependent $^1\Pi$ and $^1\Sigma^+$ character of the wavefunctions in the a$^3\Pi, v=0$ state into account.

\section{\label{sec:hyperfinev5}Hyperfine structure in the $\text{a}^\text{3}  \boldsymbol{\Pi} \text{, v = 5}$ state}
\begin{figure}[b]
		\centering
		\includegraphics[width=0.48\textwidth]{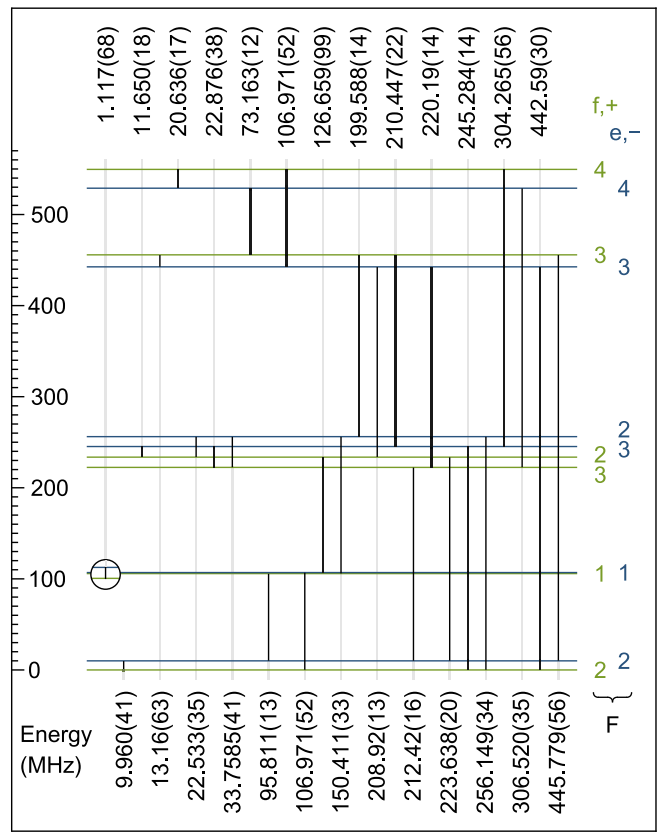}
	\caption{\label{fig:RF}Hyperfine energy level diagram of the $J=1$ level in the a$^3\Pi_1, v=5$ state. All 26 allowed rf transitions are indicated with their measured frequencies and error bars (in MHz). The quantum numbers $F$ and the parities of the levels are given.}
\end{figure}

The hyperfine structure results from the interaction of the nuclear spins of the aluminum and fluorine atoms with the electronic and rotational degrees of freedom, and is described by the term $H_{\textrm{hfs}}$ in the molecular Hamiltonian, Eq.~(\ref{eq:hamiltonian}). The vectorial coupling of the rotational angular momentum $\mathbf{J}$ with the spins of the aluminum nucleus $\mathbf{I}_\text{Al}$ and the fluorine nucleus $\mathbf{I}_\text{F}$ leads to the final vector $\mathbf{F}$, i.e. $\mathbf{F}$ = $\mathbf{J}$ + $\mathbf{I}_\text{Al}$ + $\mathbf{I}_\text{F}$. 

Here, we present our investigation of the hyperfine structure of the a$^3\Pi_1,v=5$ state. We measure all 26 allowed radio frequency transitions between $F$ levels of opposite parity within the $J=1$ level and a few transitions in the $J=2$ level. To measure the transitions in the $J=1$ level, the molecules are prepared with the PDA on the $R_2(0)$ or $Q_2(1)$ line in the six, definite parity $F$ levels. The metastable molecules subsequently pass through an rf transmission line centered at $z_{\text{rf}}=z_{\text{PDA}}+10$~cm, where they are exposed to rf radiation in the 1--500~MHz range that propagates perpendicular to the direction of the molecular beam.
Since the population in the X$^1\Sigma^+, v=5$  state is about two orders of magnitude less than that in the vibrational ground state, a relatively short transmission line of  57~mm is used in order to be able to record spectral lines with a high enough signal-to-noise.

The ambient magnetic field in the rf interaction region is about 20~\SI[parse-numbers = false]{\micro T.}. The exact timing and duration of the rf pulse is controlled by an rf switch. When the rf radiation is tuned to an electric dipole allowed transition, the population is transferred to an $F$ level with opposite parity. Given the many $M_F$ sublevels involved in the transitions and the presence of a weak magnetic field, it is not possible to apply a perfect \mbox{$\pi$ pulse}, and the duration and power of the applied rf radiation is thereby merely adjusted to maximize the total population transfer. After exiting the rf interaction region, the molecules enter the detection region, where parity-selective excitation to the b$^3\Sigma^+, v=5$ state followed by ionization takes place in a two-color (1+1$'$)~REMPI process. By recording the parent ion signal while scanning the frequency of the rf radiation, background-free rf spectra are obtained.

The widths of the observed rf spectral lines are mainly determined by Zeeman broadening due to the ambient magnetic field. Depending on the difference of the Landé $g_\text{F}$ factors of the coupled levels, the full widths at half maximum range typically from 40 to 200~kHz. The center frequency and width of in total 41 hyperfine transitions are determined with Gaussian fits. The measured frequencies for the 26 possible transitions within the $J=1$ level and their assignments in the energy level scheme are given in Fig.~\ref{fig:RF}, together with the measured error bars. All measured frequencies are fitted to the Hamiltonian, and the obtained hyperfine structure parameters are given in Table~\ref{tab:hyperfineparameter}. In the fit, several hyperfine parameters are kept fixed at the values found for the $v=0$ state. The values for the spin-orbit coupling constant and for the rotational constant of $v=5$ are taken from Sec.~\ref{sec:Rot} and are also kept fixed in this fit. The standard deviation of the fit is 20~kHz. As expected, the overall structure of the hyperfine levels in $v=5$ is very similar to that in $v=0$. 

\begin{table}

\caption{$\Lambda$-doubling parameter $q$ and hyperfine structure parameters for the a$^3\Pi, v=5$ state obtained from the best fit to the experimental data together with the product of the standard deviation and $\sqrt{Q}$ (all values in MHz). These values are compared with those determined earlier for the a$^3\Pi, v=0$ state. 
For $v=0$, the parameters 
$A_0 = 1420870$~MHz and
$\lambda_0 = 2659$~MHz
are kept fixed\cite{alf}
whereas for  $v=5$ 
$A_5 = 1431589$~MHz and
$\lambda_5 =2766$~MHz
are used.\cite{max} }
\label{tab:hyperfineparameter}
\begin{ruledtabular}
\begin{tabular}{ldddc}
	Parameter         & \multicolumn{1}{c}{Value (MHz)}   & \multicolumn{1}{c}{$\textrm{SD}\cdot\sqrt{\textrm{Q}}$}  & \multicolumn{1}{c}{Value (MHz)}   & \multicolumn{1}{c}{$\textrm{SD}\cdot\sqrt{\textrm{Q}}$} \\
	& \multicolumn{1}{c}{$v=0$} & \multicolumn{1}{c}{$v=0$} & \multicolumn{1}{c}{$v=5$} & \multicolumn{1}{c}{$v=5$} \\
    \hline	
    
$B_{\text{v}}$ & 16634.7 & 0.0010 & 15942.1 & fixed \\
$\gamma$ & -7.6089 & 0.0526 & -7.6089 & fixed \\
$o$ & 4968.32 & 0.0510 & 4968.32 & fixed \\
$o^{(R)}$ & -0.0061 & 0.0012 & -0.0061 & fixed \\
$p$ & -24.3462 & 0.0510 & -24.3462 & fixed \\
$q$ & -1.8176 & 0.0023 & -1.9192 & 0.0009 \\
$a$(Al) & 199.162 & 0.0353 & 200.2192 & 0.1339 \\
$b_F$(Al) & 1247.97 & 0.2783 & 1215.9123 & 2.9445 \\
$b_F^{(R)}$(Al) & 0.0222 & 0.0045 & 0.0222 & fixed \\
$c$(Al) & -21.0093 & 0.4124 & -21.0093 & fixed \\
$c^{(R)}$(Al) & -0.0568 & 0.0114 & -0.0568 & fixed \\
$d$(Al) & 121.908 & 0.0155 &  124.1718 &  0.0604  \\
$eq_0Q$(Al) & -12.9921 & 0.0159 &  -10.8595 & 0.0811\\
$eq_0Q_{LS}$(Al) & -0.0392 & 0.0042 & -0.0392 & fixed \\
$eq_2Q$(Al) & 51.1137 & 0.0054 &  51.8014 & 0.0220 \\
$C_I$(Al) & -0.0569 & 0.0169 & -0.0569 & fixed \\
$C'_I$(Al) & -0.0115 & 0.0007 & -0.0115 & fixed \\
$a$(F) & 207.135 & 0.0109 & 197.2087 & 0.4804 \\
$a^{(R)}$(F) & -0.0688 & 0.0038 & -0.0688 & fixed \\
$b_{F}$(F) & 169.629 & 0.1605 & 146.1175 & 10.9697 \\
$b_F^{(R)}$(F) & -0.0346 & 0.0019 & -0.0346 & fixed \\
$c$(F) & 122.804 & 0.2596 & 122.804 & fixed \\
$d$(F) & 119.277 & 0.0868 & 112.8507 &  0.3054 \\
$C'_I$(F) & 0.0346 & 0.0041 & 0.0346 & fixed \\

\end{tabular}
\end{ruledtabular}
\end{table}

For a comparison of these hyperfine parameters to theory, the electric field gradients (EFGs) and the nuclear quadrupole coupling constant $eq_0Q$, the isotropic Fermi contact coupling constants $b_F$ and the anisotropic spin dipole coupling constants $c$ and $d$ are calculated by the density functional theory (DFT) method implemented in Gaussian 2016. \cite{g16} The hybrid exchange-correlation functional CAM-B3LYP \cite{yanai2004new} with aug-cc-pV5Z basis set\cite{av5zdunning1989a,av5zkendall1992a,av5zwoon1993a} is used for the a$^3\Pi$ state of AlF. To obtain these constants, we calculate the interaction potential of AlF for 23 points ranging from 1.0 to 6.0 \AA. The nuclear quadrupole coupling constant $eq_0Q$ (in MHz) is calculated from its relationship between the EFGs (in atomic units) and the nuclear quadrupole moment $Q$ (in MBarn) as \cite{bieron2001nuclear,aerts2019revised}
\begin{equation}
\label{eq:eQq_Q_Vzz}
    eq_0Q = \frac{Q \braket{V_{zz}}_v}{4.255958}
\end{equation}
where $Q=146.6$~Mb for the $^{27}$Al atom, \cite{pyykko2018year} and $\braket{V_{zz}}_{v} = \braket{\psi_{v}| V_{zz} |\psi_{v}} $ stands for the expectation value of $V_{zz}$ for a given vibrational state $|\psi_{v}\rangle$. Bound vibrational states are calculated using a discrete variable representation (DVR) approach \cite{dvrlight1985generalized,dvrlill1982discrete} employing 200 DVR points whereas the interaction potential is fitted to the DFT potential energy curve, characterized by the values given in Table \ref{tab:AlF_DFT_constants}. The resulting hyperfine constants are summarized in Table~\ref{Tablev5}.
Note that the hyperfine parameter $a$ is not listed.
Often, the expression $a=d+c/3$ is used to deduce the value of $a$.\cite{dousmanis} However, it is clear from the experimentally determined parameters listed in Table~\ref{tab:hyperfineparameter} that this expression does not hold for the a$^3\Pi$ state of AlF.

\begin{table}
\caption{Spectroscopic parameters of AlF, including the term value $T_\text{e}$, the equilibrium internuclear distance $r_\text{e}$ and the harmonic frequency $\omega_e$ at the CAM-B3LYP/aug-cc-pV5Z level.}
\label{tab:AlF_DFT_constants}
\begin{tabular}{c|ccc}
\hline \hline
state        & $T_\text{e}$ (cm$^{-1}$) & $r_\text{e}$ (\AA) & $\omega_\text{e}$ (cm$^{-1}$) \\[0.5ex]
\hline
X$^1 \Sigma^+$ & 0                & 1.659         & 792.87         \\
a$^3 \Pi$   & 25587.86         & 1.655         & 818.72           \\
\hline \hline
\end{tabular}
\end{table}

\begin{table}[]
\small
\renewcommand{\arraystretch}{1.3}
\caption{The DFT-calculated hyperfine constants $eQq(\text{Al})$, $b_F$(Al), $b_F$(F), $c$(Al), $c$(F), $d$(Al) and $d$(F) (in MHz) of the a$^3\Pi$ state of AlF for different vibrational levels.
}
\label{Tablev5}
\begin{tabular}{c|ccccccc}
\hline \hline

$v$   & $eq_0Q$(Al) & $b_F$(Al) & $b_F$(F) & $c$(Al) & $c$(F) & $d$(Al) & $d$(F)  \\
\hline
0    &  $-$12.221  & 1186.2 &  150.1    & $-$24.6   &  154.3  & 126.5  &  128.6 \\
1    &  $-$11.763  & 1189.8 &  147.0    & $-$25.3   &  160.8  & 126.9  &  127.1 \\
2    &  $-$11.309  & 1193.1 &  144.0    & $-$25.9   &  167.6  & 127.4  &  125.5 \\
3    &  $-$10.857  & 1196.2 &  141.0    & $-$26.6   &  174.7  & 127.8  &  124.0 \\
4    &  $-$10.407  & 1199.1 &  138.0    & $-$27.2   &  182.0  & 128.3  &  122.4 \\
5    &  $-$9.957   & 1201.7 &  135.1    & $-$27.9   &  189.7  & 128.7  &  120.8 \\

\hline \hline                                                      
\end{tabular}                                                      
\end{table}           

\section{\label{sec:subnat}Absence of lifetime broadening in rf spectra of the $\text{a}^\text{3} \boldsymbol{\Pi}_\text{1} \text{, v = 0}$ state}

\begin{figure}
\centering
\begin{subfigure}[b]{0.48\textwidth}
   \includegraphics[width=1\linewidth]{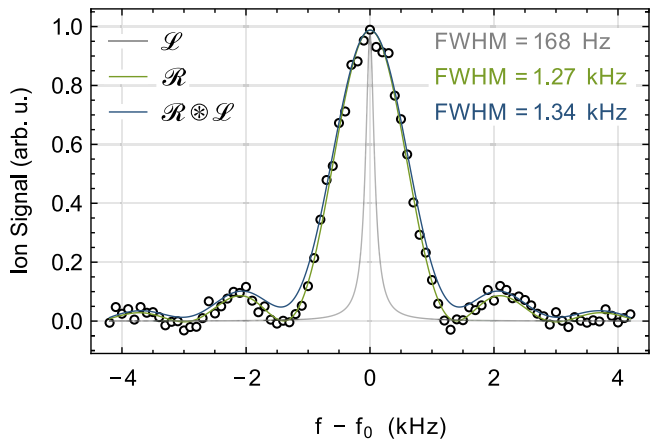}
\end{subfigure}

\caption{\label{fig:21MHz}
Lineshape of the $F=4, + \leftarrow F=4, -$ transition within the a$^3\Pi_1, v=0, J=1$ level at $f_0=21.82640$~MHz.
The circles are the experimental values. The grey curve shows a Lorentzian profile $\mathcal{L}(f)$ with a linewidth of $(\pi \tau_1)^{-1}$. 
The green and blue curves show a Rabi lineshape $\mathcal{R}(f)$ and a convolution of the Rabi lineshape with the Lorentzian $\mathcal{R}(f)\circledast\mathcal{L}(f)$.
}
\end{figure}

Spectral lines with subnatural linewidths can be measured when one selectively probes those atoms or molecules that have stayed in the excited state longer than their natural lifetime.\cite{walther}
More interestingly, when both levels involved in the transition are metastable and have the same lifetime, no additional broadening of the Rabi lineshape due to the finite lifetime of the levels is expected at all in a time-resolved measurement.\cite{burshtein1988}
The latter can be tested already for interaction times that are shorter than the lifetime, provided the Rabi-lineshape is recorded with a sufficient accuracy. 
To explicitly demonstrate this, we record the $F=4, + \leftarrow F=4, -$ transition within the a$^3\Pi_1, v=0, J=1$ level with an rf interaction time that is about one third of the lifetime. 
For this, a parallel plates transmission line of 30~cm length is used. 
The electric field propagates along the molecular beam with amplitude~$\mathcal{E}_0$. Using xenon as a carrier gas, the total interaction time of the molecules with the rf radiation field is 650~\SI[parse-numbers = false]{\micro \second.} The rf interaction region is shielded by a \SI[parse-numbers = false]{\micro \textrm{-metal}} cylinder, resulting in an ambient magnetic field of $<3$~\SI[parse-numbers = false]{\micro T} that is predominantly parallel to the electric field. The value of $\mathcal{E}_0$ is experimentally set such that the population transfer is optimized. The molecules in the $F=4, +$ level are state-selectively detected shortly after exiting the rf interaction zone using \mbox{(1+2) REMPI}. The measured ion signal as a function of the rf frequency $f$ is shown in Fig.~\ref{fig:21MHz}.

The calculated Rabi lineshape $\mathcal{R}(f)$ contains the eight possible $\Delta M_F=0$ components. The ambient magnetic field splits these eight components apart. The Landé $g_F$ factors for the upper and lower level are almost identical, with $\Delta g_{F}$ = $1.1\cdot10^{-4}$. As a result, the outermost components are separated by about 36~Hz. Transitions with $\Delta M_F=\pm1$ are shifted by $\pm4.7$~kHz but are not observed due to the near parallel orientation of the electric and magnetic fields. The green curve in Fig.~\ref{fig:21MHz} shows a fit of $\mathcal{R}(f)$ to the experimental data with the amplitude, central frequency and the value of $M_F \mu \mathcal{E}_0$ as free fit parameters. Here, $\mu$ is the effective transition dipole moment for the $M_F=1$ $\leftarrow$ $M_F=1$ transition. The fitted FWHM of the Rabi lineshape is 1.27~kHz.

To confirm that we indeed do not observe any broadening due to the finite lifetime of the two levels involved,  we fit a convolution of the Rabi lineshape with a Lorentzian $\mathcal{L}(f)$. This lineshape is shown as the blue curve in Fig.~\ref{fig:21MHz}. As the lifetime $\tau_1$ of both coupled levels is 1.89~ms (see Table~\ref{tab:LifetimeCalc}), the natural width of the Lorentzian that would be observed in a stationary measurement is given by $(\pi \tau_1)^{-1}$ as 168~Hz, shown as the grey curve in Fig.~\ref{fig:21MHz}. The pure Rabi lineshape fits the experimental data best. In particular the minima of the experimental signal next to the central peak are systematically described better when no broadening due to the lifetime of the metastable levels is included.

\section{\label{sec:conslusion}Conclusions and outlook}

\begin{table*}[t]
\centering
\caption{\label{Fcf}Calculated Franck-Condon factors for the a$^3\Pi$ -- X$^1\Sigma^+$ system of AlF.}
\begin{ruledtabular}
\begin{tabular}{rl|ccccccccccc}
&&& \multicolumn{10}{c}{a$^3\Pi$}\\
&
& $v=0$ 
& $v=1$ 
& $v=2$ 
& $v=3$ 
& $v=4$ 
& $v=5$ 
& $v=6$ 
& $v=7$ 
& $v=8$ 
& $v=9$ 
& $v=10$ \\

\hline 
\multirow{11}{*}{X$^1\Sigma^+$}
&$ v=0$ & 0.992 &  7.5e$-$09 &  5.8e$-$05 & 4.6e$-$09 & 8.8e$-$09 & 2.3e$-$10 &  1.2e$-$10 & 1.0e$-$10 & 2.1e$-$11 & 1.9e$-$12&  3.4e$-$14 \\ 
&$ v=1$ & 0.008 &    0.976 &   0.016  &2.0e$-$04&  5.8e$-$08& 2.3e$-$08 & 5.4e$-$10 & 6.0e$-$10 & 6.2e$-$10 & 1.4e$-$10 & 1.4e$-$11 \\ 
&$ v=2$ & 1.2e$-$05 &   0.016 &    0.958  &  0.025 & 4.0e$-$04& 2.8e$-$07 & 3.4e$-$08 & 3.7e$-$10 & 1.7e$-$09 & 2.0e$-$09 & 5.3e$-$10 \\ 
&$ v=3$ & 3.0e$-$07 & 4.7e$-$05 &   0.025 &    0.938 &   0.035 & 7.4e$-$04 & 8.8e$-$07 & 3.9e$-$08 & 3.9e$-$11 & 3.5e$-$09 & 5.2e$-$09 \\ 
&$ v=4$ & 5.9e$-$08 & 9.6e$-$07 & 1.2e$-$04 &   0.036  &   0.916 &   0.046 &  0.001 & 2.3e$-$06 & 4.7e$-$08 & 4.1e$-$09 & 5.9e$-$09 \\ 
&$ v=5$ & 4.4e$-$09 & 2.2e$-$07 & 1.8e$-$06 &  2.5e$-$04  &  0.047   &  0.891 &   0.058 &  0.002  & 5.4e$-$06 & 7.5e$-$08 & 2.7e$-$08 \\ 
&$ v=6$ & 1.6e$-$11 & 1.7e$-$08 & 5.2e$-$07 & 2.4e$-$06  &4.8e$-$04   &  0.060   &   0.865  &  0.072  & 0.003&  1.2e$-$05 & 1.8e$-$07 \\ 
&$ v=7$ & 1.3e$-$10 & 4.7e$-$11  &  3.7e$-$08 & 9.9e$-$07 & 2.3e$-$06 & 8.4e$-$04  &  0.074   &  0.835 &   0.086  & 0.003 & 2.6e$-$05 \\ 
&$ v=8$ & 9.3e$-$11 & 7.4e$-$10 & 4.2e$-$11&  6.3e$-$08&  1.6e$-$06 & 1.3e$-$06  & 0.001 &   0.089 &    0.803   &   0.101  & 0.005 \\ 
&$ v=9$ & 2.4e$-$11 & 6.0e$-$10 & 2.3e$-$09 & 2.8e$-$13  &9.2e$-$08&  2.5e$-$06  &1.5e$-$07  & 0.002   & 0.105  &   0.768  &   0.119 \\ 
&$ v=10$ & 3.0e$-$12 & 1.8e$-$10&  2.1e$-$09 & 5.7e$-$09 & 2.8e$-$10 & 1.3e$-$07  &3.4e$-$06 & 7.3e$-$07 &  0.003   &  0.122    & 0.728 \\ 
\end{tabular}
\end{ruledtabular}
\end{table*}

This paper reports on the spectroscopic characterization of the a$^3\Pi$ state of AlF, using a jet-cooled, pulsed molecular beam machine.
The ro-vibrational constants determined in this work for the a$^3\Pi$ state enable the construction of an accurate electronic potential. Together with the best known potential for the X$^1\Sigma^+$ state, this is used to calculate the Franck-Condon factors for the a$^3\Pi$ -- X$^1\Sigma^+$ system as given in Table~\ref{Fcf}. The system is highly diagonal, the transitions from the a$^3\Pi, v=0$ state to the
X$^1\Sigma^+,v=1$ and $v=2$ states are more than two and almost five orders of magnitude weaker than those to the $v=0$ state.

The lifetimes of the rotational levels in the a$^3\Pi$ state of AlF depend strongly on the $\Omega$ manifold and on their $e/f$ character. 
From the $N=1$ level in the X$^1\Sigma^+, v=0$ state, rotationally closed transitions can be induced to
(i) the $J=1, f$ level in the A$^1\Pi, v=0$ state,
(ii) the $J=1, f$ levels in the a$^3\Pi_1, v=0$ state as well as the
$J=1, f$ levels in the a$^3\Pi_0, v=0$ state and
(iii) the $J=0, e$ level in the a$^3\Pi_0, v=0$ state.
All these transitions are vibrationally closed to better than 99.2~\%, and the lifetimes of the upper levels vary from 1.9~ns ($J=1, f$ in A$^1\Pi$) via 1.89~ms ($J=1, f$ in a$^3\Pi_1$) and 25~ms ($J=0, e$ in a$^3\Pi_0$) to 3.4~s ($J=1, f$ in a$^3\Pi_0$). 
The latter value is calculated from the fraction of $\Omega=1$ character in the wavefunction of the $J=1$, $f$ level in a$^3\Pi_0$ and the measured lifetime of the $J=1$, $f$ level in a$^3\Pi_1$.

The information on the Franck-Condon matrix and the lifetime of the rotational levels of the a$^3\Pi$ -- X$^1\Sigma^+$ system obtained here is crucial for future laser cooling experiments. The duration that molecules stay in a certain rotational level in the a$^3\Pi, v=0$ state can be made considerably shorter than their lifetime by driving the A$^1\Pi, v=0$ $\leftarrow$  a$^3\Pi, v=0$ transition.
From the A$^1\Pi, v=0$ state, the molecules decay rapidly to the ground state.
In this way, the scattering rate on the narrow a$^3\Pi, v=0$ $\leftarrow$ X$^1\Sigma^+, v=0$ transition can be increased.
Four transitions are required to return to the initial $N=1$ rotational level in the X$^1\Sigma^+, v=0$ state: 
(i) excitation on the $P_1(1)$ line of the a$^3\Pi_0, v=0$ $\leftarrow$ X$^1\Sigma^+, v=0$ band, 
(ii)  rf transitions from the $e$ to the $f$ levels in a$^3\Pi_0, v=0, J=0$ level,
(iii) excitation on the $R_1(0)$ line of the A$^1\Pi, v=0$ $\leftarrow$ a$^3\Pi_0, v=0$ band and
(iv) spontaneous emission on the $Q(1)$ line of the A$^1\Pi, v=0$ $\rightarrow$ X$^1\Sigma^+, v=0$ band.

The outcome of the quantum chemistry calculations given in Table~\ref{Tablev5} are compared to the experimentally determined values of the hyperfine parameters listed in Table~\ref{tab:hyperfineparameter}. The sign and magnitude of the calculated hyperfine parameters is seen to be in good agreement with the experimentally determined ones, with deviations mostly below 5~\%, although somewhat larger for the $c$ parameters. 
The relative change of the hyperfine parameters in going from $v=0$ to $v=5$ is also correctly reproduced by the calculations. For the $b_F$ parameters, defined as $b_F$ = $b$ + $c$/3, the latter is less good. This deviation results from the need to keep the $c$ parameter fixed in the fit. The changes of the hyperfine parameters in going from $v=0$ to $v=5$ result predominantly from the concomitant increase of the internuclear distance. The spectroscopic parameters of the a$^3\Pi$ state, in particular the hyperfine structure constants and their variation with the vibrational quantum number, are ideal benchmarks for quantum chemistry calculations.

Although not commonly realised, spectral lines without any broadening due to the finite lifetime of the levels can be obtained in a time-resolved measurement, when the coupled levels have the same lifetime. By recording an rf transition between $\Lambda$-doublet components in the a$^3\Pi_1$ state, we have been able to confirm this. Using an even longer transmission line, thereby increasing the interaction time, this effect can be exploited further.

\begin{acknowledgments}
We thank Marco De~Pas, Uwe Hoppe, Sebastian Kray, Christian Schewe and Klaus-Peter Vogelgesang for excellent technical support and Sid Wright for critical reading of the manuscript. N.~W. acknowledges support by the International Max Planck Research School for Elementary Processes in Physical Chemistry. This project has received funding from the European Research Council (ERC) under the European Union’s Horizon 2020 research and innovation programme (CoMoFun, Grant Agreement No. 949119).
\end{acknowledgments}

\section{Data Availability Statement}
The data that support the findings of this study are available from the corresponding author upon reasonable request

\section*{Author Contributions}
N.~W. carried out the data acquisition, analyzed and visualized the data and wrote the manuscript. 
M.~D., S.~M., J.~S.and S.~T. took on parts of the data acquisition.
B.~S. analyzed the data, X.~L. and J.~P carried out quantum chemistry calculations.
G.~M. supervised and conceptualized the project and edited the manuscript.

\appendix

\section{Observed and calculated transition frequencies}\label{ap:RotLineList}

Table~\ref{tab:RotTable} lists the 30 measured rotational transition frequencies of the  a$^3\Pi, v$ $\leftarrow$ X$^1\Sigma^+, v$ bands.
In the first column, the experimentally determined transition frequency $\tilde{\nu}_\text{exp}$ is given together with the statistical uncertainty from the Gaussian fit.
The systematic experimental uncertainty is estimated to be at maximum 200 MHz, mainly determined by the intrinsic laser profile and the uncertainty of the SBS shift. In the second column, the difference between $\tilde{\nu}_\text{exp}$ and the calculated frequency $\tilde{\nu}_\text{calc}$ is given. The next columns list 
the vibrational quantum number $v$, the rotational quantum number $J$, 
the parity $p$ and, if applicable, the quantum number $\Omega$ of the levels involved in the transition. The indices define whether the quantum numbers belong to the a$^3\Pi$ or to the X$^1\Sigma^+$ state. Only transitions with $\Delta v = 0$ are measured. The standard deviation of the fit is 38~MHz.

Table~\ref{tab:RFTable} lists the 41 measured rf transition frequencies in the a$^3\Pi_1, v=5$ state. In the first column, the experimentally determined transition frequency $f_\text{exp}$ is given together with the statistical uncertainty from the Gaussian fit. In the second column, the difference between $f_\text{exp}$ and the calculated frequency $f_\text{calc}$ is given. The next columns list 
the quantum numbers $J$, $F$ and the parity $p$ of the levels involved in the transition. The index $n$ is introduced to distinguish levels with the same values of $F$ and $p$, in order of increasing energy. The lowest energy level for a given combination $F,p$ in the a$^3\Pi$ state gets the index $n=1$. From the 26 possible transitions in the $J=1$ level, two overlap within $< 50$~kHz and could not be resolved experimentally and are listed with the same value of $f_\text{exp}$. The standard deviation of the fit is 20~kHz.

\begin{table}
\caption{\label{tab:RotTable}Observed and calculated transition frequencies of the  a$^3\Pi, v$ $\leftarrow$ X$^1\Sigma^+, v$ bands. All values in cm$^{-1}$.}
\begin{ruledtabular}
\begin{tabularx}{1\textwidth}{lrrrrrrrrr}
 $~~~~~~~\tilde{\nu}_\text{exp}$ &  $\tilde{\nu}_\text{exp}-\tilde{\nu}_\text{calc}$ & $v_{\text{a}}$ & $\Omega_{\text{a}}$   & $J_{\text{a}}$  & $p_{\text{a}}$  & $v_{\text{X}}$  & $J_{\text{X}}$  & $p_{\text{X}}$ \\ [0.5ex]
\hline 
27205.55305(28) & 0.0013 & 0 & 0 & 0 & + & 0 & 1 & -- \\
27233.82971(49) & 0.0006 & 1 & 0 & 0 & + & 1 & 1 & -- \\
27254.149(29) & $-$0.0020 & 0 & 1 & 4 & -- & 0 & 4 & + \\
27254.196(30) & 0.0011 & 0 & 1 & 5 & + & 0 & 5 & -- \\
27255.17237(60) & 0.0000 & 0 & 1 & 1 & -- & 0 & 0 & + \\
27262.45333(63) & 0.0007 & 2 & 0 & 0 & + & 2 & 1 & -- \\
27282.544(30) & 0.0035 & 1 & 1 & 5 & + & 1 & 5 & -- \\
27282.595(30) & $-$0.0005 & 1 & 1 & 6 & -- & 1 & 6 & + \\
27283.50261(75) & 0.0002 & 1 & 1 & 1 & -- & 1 & 0 & + \\
27291.37627(64) & 0.0013 & 3 & 0 & 0 & + & 3 & 1 & -- \\
27311.182(29) & $-$0.0023 & 2 & 1 & 4 & -- & 2 & 4 & + \\
27311.231(29) & $-$0.0013 & 2 & 1 & 5 & + & 2 & 5 & -- \\
27311.357(30) & $-$0.0002 & 2 & 1 & 7 & + & 2 & 7 & -- \\
27312.17743(80) & 0.0001 & 2 & 1 & 1 & -- & 2 & 0 & + \\
27340.130(30) & 0.0001 & 3 & 1 & 3 & + & 3 & 3 & -- \\
27341.14959(50) & $-$0.0003 & 3 & 1 & 1 & -- & 3 & 0 & + \\
27349.91730(53) & 0.0001 & 5 & 0 & 0 & + & 5 & 1 & -- \\
27369.406(29) & $-$0.0004 & 4 & 1 & 4 & -- & 4 & 4 & + \\
27369.460(29) & 0.0007 & 4 & 1 & 5 & + & 4 & 5 & -- \\
27369.524(30) & 0.0004 & 4 & 1 & 6 & -- & 4 & 6 & + \\
27370.37158(65) & 0.0008 & 4 & 1 & 1 & -- & 4 & 0 & + \\
27398.842(30) & 0.0028 & 5 & 1 & 4 & -- & 5 & 4 & + \\
27398.895(30) & $-$0.0004 & 5 & 1 & 5 & + & 5 & 5 & -- \\
27398.964(30) & 0.0017 & 5 & 1 & 6 & -- & 5 & 6 & + \\
27399.78838(80) & $-$0.0005 & 5 & 1 & 1 & -- & 5 & 0 & + \\
27428.414(30) & $-$0.0019 & 6 & 1 & 4 & -- & 6 & 4 & + \\
27428.545(30) & $-$0.0014 & 6 & 1 & 6 & -- & 6 & 6 & + \\
27429.3513(13) & 0.0001 & 6 & 1 & 1 & -- & 6 & 0 & + \\
27458.035(30) & 0.0013 & 7 & 1 & 3 & + & 7 & 3 & -- \\
27459.0033(28) & 0.0004 & 7 & 1 & 1 & -- & 7 & 0 & + \\
\end{tabularx}
\end{ruledtabular}
\end{table}

\begin{table}
\caption{\label{tab:RFTable}Observed and calculated rf transition frequencies in the a$^3\Pi_1, v=5$ state. All values in MHz.
}
\begin{ruledtabular}
\begin{tabularx}{1\textwidth}{lrcccccccc} 
 $~~~~~~~f_\text{exp}$ &
 $f_\text{exp}-f_\text{calc}$ &
  $J'$  & 
  $F'$  & 
  $p'$  & 
  $n'$  & 
  $J$  & 
  $F$  & 
    $p$  & 
  $n$ \\ [0.5ex]
  
\hline 
1.117(68) & $-$0.047 & 1 & 1 & -- & 1 & 1 & 1 & + & 1 \\
9.960(41) & $-$0.031 & 1 & 2 & -- & 2 & 1 & 2 & + & 2 \\
11.650(18) & 0.004 & 1 & 3 & -- & 3 & 1 & 2 & + & 2 \\
13.16(63) & $-$0.015 & 1 & 3 & + & 3 & 1 & 3 & -- & 3 \\
20.636(17) & $-$0.010 & 1 & 4 & + & 4 & 1 & 4 & -- & 4 \\
22.533(35) & 0.027 & 1 & 2 & -- & 2 & 1 & 2 & + & 2 \\
22.876(38) & $-$0.025 & 1 & 3 & -- & 3 & 1 & 3 & + & 3 \\
33.7585(41) & $-$0.003 & 1 & 2 & -- & 2 & 1 & 3 & + & 3 \\
73.163(12) & $-$0.007 & 1 & 4 & -- & 4 & 1 & 3 & + & 3 \\
95.811(13) & 0.009 & 1 & 1 & + & 1 & 1 & 2 & -- & 2 \\
106.971(52) & 0.014 & 1 & 1 & -- & 1 & 1 & 2 & + & 2 \\
106.971(52) & $-$0.023 & 1 & 4 & + & 4 & 1 & 3 & -- & 3 \\
126.659(99) & $-$0.023 & 1 & 2 & + & 2 & 1 & 1 & -- & 1 \\
150.411(33) & 0.058 & 1 & 2 & -- & 2 & 1 & 1 & + & 1 \\
199.588(14) & $-$0.004 & 1 & 3 & + & 3 & 1 & 2 & -- & 2 \\
208.92(13) & $-$0.005 & 1 & 3 & -- & 3 & 1 & 2 & + & 2 \\
210.447(22) & $-$0.005 & 1 & 3 & + & 3 & 1 & 3 & -- & 3 \\
212.42(16) & 0.022 & 1 & 3 & + & 3 & 1 & 2 & -- & 2 \\
220.19(14) & 0.014 & 1 & 3 & -- & 3 & 1 & 3 & + & 3 \\
223.638(20) & $-$0.011 & 1 & 2 & + & 2 & 1 & 2 & -- & 2 \\
245.284(14) & $-$0.001 & 1 & 3 & -- & 3 & 1 & 2 & + & 2 \\
256.149(34) & 0.004 & 1 & 2 & -- & 2 & 1 & 2 & + & 2 \\
304.265(56) & $-$0.005 & 1 & 4 & + & 4 & 1 & 3 & -- & 3 \\
306.520(35) & $-$0.003 & 1 & 4 & -- & 4 & 1 & 3 & + & 3 \\
442.59(30) & 0.026 & 1 & 3 & -- & 3 & 1 & 2 & + & 2 \\
445.779(56) & 0.033 & 1 & 3 & + & 3 & 1 & 2 & -- & 2 \\
17.493(26) & 0.004 & 2 & 3 & + & 16 & 2 & 3 & -- & 16 \\
22.047(35) & 0.005 & 2 & 2 & -- & 13 & 2 & 1 & + & 8 \\
22.490(25) & $-$0.027 & 2 & 4 & -- & 15 & 2 & 4 & + & 15 \\
36.283(11) & $-$0.001 & 2 & 5 & -- & 13 & 2 & 5 & + & 13 \\
47.681(15) & 0.010 & 2 & 3 & -- & 15 & 2 & 2 & + & 13 \\
50.892(17) & $-$0.012 & 2 & 3 & -- & 16 & 2 & 2 & + & 14 \\
85.771(13) & 0.000 & 2 & 4 & -- & 14 & 2 & 3 & + & 15 \\
87.800(19) & 0.001 & 2 & 0 & + & 3 & 2 & 1 & -- & 8 \\
90.010(11) & 0.015 & 2 & 5 & -- & 13 & 2 & 4 & + & 15 \\
93.064(58) & $-$0.032 & 2 & 3 & + & 16 & 2 & 2 & -- & 14 \\
94.513(27) & $-$0.005 & 2 & 4 & + & 15 & 2 & 3 & -- & 16 \\
99.515(21) & $-$0.033 & 2 & 4 & -- & 15 & 2 & 3 & + & 16 \\
109.408(16) & 0.012 & 2 & 3 & + & 15 & 2 & 2 & -- & 13 \\
119.663(26) & 0.002 & 2 & 5 & + & 13 & 2 & 4 & -- & 14 \\
147.046(28) & $-$0.002 & 2 & 5 & -- & 13 & 2 & 4 & + & 14 \\
\end{tabularx}
\end{ruledtabular}
\end{table}
\FloatBarrier


\section{Radiative lifetimes in the $\text{a}^\text{3} \boldsymbol{\Pi} \text{, v = 0}$ state}\label{ap:lifetimeaDerivation}
The wavefunctions $\ket{\psi_{\Omega'}(J)}$ for the $J$ levels in the a$^3\Pi_{\Omega'}$ manifolds can be written as
\begin{align}
\begin{aligned}
    \ket{\psi_{\Omega'}(J)} = 
    c_{\Omega',0}&(J) \ket{\psi_0(J)}  \,+\\
     &c_{\Omega',1}(J) \ket{\psi_1(J)} \,+ \, 
      c_{\Omega',2}(J) \ket{\psi_2(J)} 
\end{aligned}
\end{align}
with the pure $\Omega$-wavefunctions $\ket{\psi_{\Omega}(J)}$ and the mixing coefficients $c_{\Omega',\Omega}(J)$. The calculated values of $c_{\Omega',\Omega}(J)$ for $J \leq 30$ are listed in  Table~\ref{tab:cfactors}. The squares of the mixing coefficients give the amount of $\Omega$ character of a given $J$ level in the a$^3\Pi_{\Omega'}$ manifold and these are normalized such that 
\begin{equation}
|c_{\Omega',0}(J)|^2+
|c_{\Omega',1}(J)|^2+
|c_{\Omega',2}(J)|^2
=1
\end{equation}

The a$^3\Pi$ state interacts with pure singlet states via spin-orbit coupling. 
For this interaction, the $\Delta \Omega=0$ selection rule holds. This implies that the strength of the interaction of a given \mbox{$J$ level} with a $^1\Sigma$, a $^1\Pi$ or a $^1\Delta$ state depends on the coefficients $c_{\Omega',0}(J)$, $c_{\Omega',1}(J)$ and $c_{\Omega',2}(J)$, respectively. 
For the a$^3\Pi$ state of AlF, both mixing with $^1\Sigma^+$ and $^1\Pi$ states is observed. The transition dipole moment $\mu_{\text{a,X}}(\Omega',J)$ for transitions from a given a$^3\Pi_{\Omega'}$, $J$ level to the X$^1\Sigma^+$ electronic ground state can be expressed as
\begin{equation}
\mu_{\text{a,X}}(\Omega',J) \propto \kappa \, c_{\Omega',0}(J) \, \text{HL}_{\Sigma}(J) +c_{\Omega',1}(J) \, \text{HL}_{\Pi}(J)
\end{equation}
Here, $\text{HL}_{\Sigma}(J)$ and $\text{HL}_{\Pi}(J)$ are the amplitudes of the normalized H\"onl-London factors for a
$^1\Sigma^+ - ^1\Sigma^+$ and for a
$^1\Pi - ^1\Sigma^+$  
transition, respectively. The parameter $\kappa$ is used to quantify the contribution from coupling with a $^1\Sigma^+$ state relative to that with a $^1\Pi$ state; these separate contributions can either interfere constructively ($\kappa>0$) or destructively ($\kappa<0$). 
The amplitudes of the normalized H\"onl-London factors $\text{HL}_{\Pi}(J)$ for $R(J-1)$, $Q(J)$ and $P(J+1)$ transitions are given by
\begin{equation}
\sqrt{\frac{J+1}{2(2J+1)}}, \quad
-\frac{1}{\sqrt{2}} 
 \quad \text{and}  \quad
 \sqrt{\frac{J}{2(2J+1)}}
\end{equation}
respectively. 
The corresponding normalized amplitudes $\text{HL}_{\Sigma}(J)$ are
\begin{equation}
\sqrt{\frac{J}{2J+1}}, \quad
0
 \quad \text{and}  \quad
-\sqrt{\frac{J+1}{2J+1}}
\end{equation}
The ratio of the integrated intensities of the $R_2(0)$ and $R_1(0)$ lines is thus given by
\begin{equation}
\label{eq:ratiolifetime}
\frac{I_{R_2(0)}}{I_{R_1(0)}}=
\frac{\left(\kappa \, c_{1,0}(1)+c_{1,1}(1)\right)^2}
{\left(\kappa \, c_{0,0}(1)+c_{0,1}(1)\right)^2}
\end{equation}

The $J=1$, $e$ levels in the a$^3\Pi_{\Omega'}$ manifolds can decay via electric dipole transitions to either the $N=0$ or the $N=2$ level in the X$^1\Sigma^+$ state, thus via the $R_{\Omega'+1}(0)$ or 
$P_{\Omega'+1}(2)$ line. 
The total decay rates of these levels for $\Omega'=0$ and $1$ are referred to as $\gamma_0$ and $\gamma_1$ and their lifetimes as $\tau_0$ and $\tau_1$, respectively.
The ratio of $\tau_0$ and $\tau_1$ is given by
\begin{align}
\begin{aligned}
\label{eq:ratiotau}
\frac{\tau_0}{\tau_1}&=
\frac{\gamma_1}{\gamma_0}\approx
\frac{\tilde{\nu}^3_{{R}_2(0)}I_{{R}_2(0)}+\tilde{\nu}^3_{{P}_2(2)}I_{{P}_2(2)}}
{\tilde{\nu}^3_{{R}_1(0)}I_{{R}_1(0)}+{\tilde{\nu}^3_{{P}_1(2)}I_{{P}_1(2)}}}
\\
&\approx
\frac{
\left(
\kappa \, c_{1,0}(1)+ c_{1,1}(1) 
\right)^2
+ 2 
\left(
- \kappa \, c_{1,0}(1) + 0.5 \, c_{1,1}(1)
\right)^2
}{
\left(
\kappa \, c_{0,0}(1)+ c_{0,1}(1) 
\right)^2
+ 2 
\left(
- \kappa \, c_{0,0}(1) + 0.5 \, c_{0,1}(1)
\right)^2
}
\end{aligned}
\end{align}
where $\tilde{\nu}$ is the transition frequency and $I$ the transition intensity of the line given by the respective index. 
In the first approximation of Eq.~(\ref{eq:ratiolifetime}), it is assumed that the Franck-Condon factor for the a$^3\Pi, v=0$ --
X$^1\Sigma^+, v=0$ transition is exactly 1.0 and in the second approximation, the frequencies of the various transitions are taken to be equal.  

The observed intensity ratio $I_{R_2(0)}/I_{R_1(0)} = 20.9 \pm 1.0$ is consistent with both 
$\kappa_+=0.243 \pm 0.013$ and $\kappa_-=-0.194 \pm 0.011$, 
resulting in ratios of 
$\tau_0$/$\tau_1=8.4 \pm 0.7$ and $\tau_0$/$\tau_1= 13.2 \pm 1.1$, respectively.
The observation that the $P_1$ lines are considerably weaker than the $R_1$ lines\cite{alf} is only consistent with the case of destructive interference, i.e. $\kappa=\kappa_-$.

\begin{table*}
\caption{\label{tab:cfactors}Calculated mixing coefficients for the wavefunctions of the $e$ and $f$ levels in the $\Omega$ manifolds of the a$^3\Pi, v=0$ state for various values of $J$.}
 \begin{ruledtabular}
 
\begin{tabular}{c|rrr|rrr|rrr}
$J$ &
$c_{0,0}$ & 
$c_{0,1}$ &
$c_{0,2}$ &
$c_{1,0}$ &
$c_{1,1}$ &
$c_{1,2}$ &
$c_{2,0}$ &
$c_{2,1}$ &
$c_{2,2}$\\ [0.5ex]

\hline
0 & 1.00000 & 0.00000 & 0.00000 &  &  &  &  &  &  \\
1 & 0.99972 & $-$0.02348 & 0.00000 & 0.02348 & 0.99972 & 0.00000 &  &  &  \\
2 & 0.99917 & $-$0.04063 & 0.00069 & 0.04063 & 0.99859 & $-$0.03428 & 0.00071 & 0.03428 & 0.99941 \\
3 & 0.99835 & $-$0.05737 & 0.00153 & 0.05736 & 0.99689 & $-$0.05411 & 0.00158 & 0.05411 & 0.99853 \\
4 & 0.99726 & $-$0.07390 & 0.00264 & 0.07389 & 0.99463 & $-$0.07243 & 0.00273 & 0.07242 & 0.99737 \\
5 & 0.99591 & $-$0.09026 & 0.00402 & 0.09026 & 0.99184 & $-$0.09007 & 0.00415 & 0.09006 & 0.99593 \\
6 & 0.99430 & $-$0.10646 & 0.00565 & 0.10645 & 0.98851 & $-$0.10728 & 0.00584 & 0.10727 & 0.99421 \\
7 & 0.99244 & $-$0.12247 & 0.00754 & 0.12245 & 0.98468 & $-$0.12415 & 0.00778 & 0.12414 & 0.99223 \\
8 & 0.99035 & $-$0.13828 & 0.00967 & 0.13826 & 0.98035 & $-$0.14071 & 0.00998 & 0.14069 & 0.99000 \\
9 & 0.98802 & $-$0.15387 & 0.01203 & 0.15384 & 0.97555 & $-$0.15697 & 0.01241 & 0.15694 & 0.98753 \\
10 & 0.98547 & $-$0.16923 & 0.01462 & 0.16919 & 0.97029 & $-$0.17293 & 0.01508 & 0.17289 & 0.98483 \\
15 & 0.96980 & $-$0.24197 & 0.03052 & 0.24186 & 0.93810 & $-$0.24794 & 0.03137 & 0.24783 & 0.96829 \\
20 & 0.95039 & $-$0.30697 & 0.05018 & 0.30677 & 0.89839 & $-$0.31429 & 0.05140 & 0.31410 & 0.94800 \\
25 & 0.92874 & $-$0.36366 & 0.07207 & 0.36336 & 0.85431 & $-$0.37164 & 0.07358 & 0.37135 & 0.92557 \\
30 & 0.90612 & $-$0.41223 & 0.09490 & 0.41184 & 0.80850 & $-$0.42036 & 0.09656 & 0.41998 & 0.90238 \\

\end{tabular}

\end{ruledtabular}
\end{table*}

\begin{table}
\caption{\label{tab:LifetimeCalc}Calculated lifetimes for the $e$ and $f$ levels in the $\Omega$ manifolds of the a$^3\Pi, v=0$ state for various values of $J$, assuming only electric dipole allowed transitions. All values in ms.}
 \begin{ruledtabular}
 
\begin{tabular}{c|cccccc}
$J$ &
a$^3\Pi_0$, $e$ & 
a$^3\Pi_0$, $f$ &
a$^3\Pi_1$, $e$ &
a$^3\Pi_1$, $f$ &
a$^3\Pi_2$, $e$ &
a$^3\Pi_2$, $f$\\ [0.5ex]

\hline

0 & 25.0 & $\infty$ &  &  &  &  \\
1 & 24.8 &  3420 & 1.89 & 1.89 &  &  \\
2 & 24.5 &     1140 & 1.89 & 1.89 &    1600 &   1600 \\
3 & 24.0 & 573 & 1.90 & 1.90 & 644 & 644 \\
4 & 23.4 & 345 & 1.91 & 1.91 & 359 & 359 \\
5 & 22.7 & 231 & 1.92 & 1.92 & 232 & 232 \\
6 & 22.0 & 166 & 1.93 & 1.93 & 164 & 164 \\
7 & 21.1 & 126 & 1.94 & 1.94 & 122 & 122 \\
8 & 20.3 & 98.6 & 1.96 & 1.96 & 95.2 & 95.3 \\
9 & 19.4 & 79.6 & 1.98 & 1.98 & 76.5 & 76.6 \\
10 & 18.5 & 65.8 & 2.00 & 2.00 & 63.0 & 63.1 \\
15 & 14.6 & 32.2 & 2.13 & 2.14 & 30.7 & 30.7 \\
20 & 11.6 & 20.0 & 2.32 & 2.34 & 19.1 & 19.1 \\
25 & 9.56 & 14.3 & 2.55 & 2.58 & 13.6 & 13.7 \\
30 & 8.13 & 11.1 & 2.83 & 2.88 & 10.6 & 10.7 \\

\end{tabular}

\end{ruledtabular}
\end{table}

\bibliography{aipsamp}

\end{document}